\documentclass[a4paper,11pt]{article}
 
 \usepackage{microtype}
\usepackage[utf8]{inputenc}
\usepackage[T1]{fontenc}
\usepackage{lmodern}
\usepackage{amsmath,amsthm,amsfonts,amssymb,bm, bbm}
\usepackage{dsfont}			% indicator function, \mathds{1}
\usepackage{enumitem}
\usepackage{graphicx}
\usepackage{booktabs}
\usepackage[authoryear,comma]{natbib}
\usepackage[right=2cm,left=2cm,top=2cm,bottom=3cm]{geometry}
\usepackage[font={small,it}]{caption}	% the options makes the font italic in floats
\usepackage[colorlinks,linkcolor=blue,citecolor=blue,urlcolor=blue]{hyperref}
\usepackage{multirow}
\usepackage{subfig}
\usepackage{quoting}
\quotingsetup{font=it}
\usepackage{url}
\usepackage{tikz}       
\usepackage{verbatim}  %draw bayesian nw
\usetikzlibrary{bayesnet}     
\usepackage{authblk}

\usepackage{forest}
\usetikzlibrary{arrows.meta, shapes.geometric, calc, shadows}

\colorlet{grey}{black!20}
\colorlet{black}{black}
\colorlet{white}{white}

\usepackage{DejaVuSansMono}
\usepackage[scaled=0.8]{beramono}

 \usepackage{array}
 \newcolumntype{L}[1]{>{\raggedright\let\newline\\\arraybackslash\hspace{0pt}}m{#1}}
\newcolumntype{C}[1]{>{\centering\let\newline\\\arraybackslash\hspace{0pt}}m{#1}}
 \newcolumntype{R}[1]{>{\raggedleft\let\newline\\\arraybackslash\hspace{0pt}}m{#1}}

\newcommand{\Y}{\mathbf{Y}}

\newcommand{\arch}{y_{ij}^{(k)} }
\newcommand{\ak}{\alpha^{(k)} }
\newcommand{\bk}{\beta^{(k)} }
\newcommand{\ti}{\theta_i }

\newcommand{\tik}{\theta_i^{(k)} }
\newcommand{\gjk}{\gamma_j^{(k)} }
\title{Modelling heterogeneity in Latent Space Models for Multidimensional Networks }

\author{Silvia D'Angelo , Marco Alfò, Thomas Brendan Murphy}

\begin{document}
\maketitle

\begin{abstract}
Multidimensional network data can have different levels of complexity, as nodes may be characterized by heterogeneous individual-specific features, which may vary across the networks. This paper introduces a class of models for multidimensional network data, where different levels of heterogeneity within and between networks can be considered. The proposed framework is developed in the family of latent space models, and it aims to distinguish symmetric relations between the nodes and node-specific features. Model parameters are estimated via a Markov Chain Monte Carlo algorithm. Simulated data and an application to a real example, on fruits import/export data, are used to illustrate and comment on the performance of the proposed models.
\end{abstract}
%-----------------------------------------------------------------------------

\section{Introduction}
\label{cap:cap3:intro}
Relational data can be, and often are, represented in the form of networks. In an observed network, dyadic relations of interest are coded as edges between nodes. When multiple relations are recorded among the same group of nodes, a multidimensional network (multiplex) is observed. If the same relation is observed through time on the same set of nodes, a dynamic network can be defined. Observed network data, either uni-dimensional or multidimensional, can exhibit different features, and these may directly influence their structure.
A frequently studied one is \emph{transitivity}, which refers to transitive relations. Roughly speaking, transitivity in social networks can be described by the ``a friend of my friend is my friend'' phenomenon. 
A popular way to model such a feature is trough latent space models, first introduced by \cite{hoff}. The basic idea is to represent the nodes in a low-dimensional unobserved space, postulating that the probability of observing an edge in the network depend on the positions of the nodes in such a space; closer objects in the latent space have an higher propensity to be connected in the network. Different latent space approaches have been proposed in the literature, either based on metrics, see \cite{hoff}, \cite{hoffbili}, or on ultrametrics, see \cite{ultrametric}. Among metric latent spaces, models based on the Euclidean distance are likely the most widespread, as they produce easily interpretable representations of the networks while being flexible enough to describe a large variety of network data. Alternatives to distance-based latent space models are the projection, \cite{hoff}, and the multiplicative latent space model, \cite{hoffbili}. Both postulate that the edge probabilities depend on the inner products of the latent coordinates. Such models address transitivity differently from distance models, as they incorporate in the latent space representation also the ``direction'' of the relation between the nodes, see \cite{hoff}. Extensions of the latent space models to multidimensional networks are described in \cite{gollini}, \cite{salter}, \cite{hoff_multi} and \cite{mio}. \cite{sewell1} introduce a latent space model for dynamic binary networks, later extended to include weighted dynamic networks, see \cite{sewell2}. \cite{durantela} developed a framework based on a dynamic latent space model to describe dynamic networks of face-to-face individual contacts. It is worth to point out that multiplex data could be also analyzed from the point of view of tensor data analysis. Within this framework, we refer to \cite{HoffT}, \cite{Bailey2014} and \cite{Zhou2015}.\\
Another interesting feature of network data is that of \emph{degree heterogeneity}, which refers to the (higher/lower) propensity of nodes to send/receive edges when compared to others. \cite{p1} proposed the so called ``$p_1$'' model, where node-specific sender and receiver effects are treated as random effects. \cite{p2} developed the ``$p_2$'' model, an extension of the ``$p_1$'' model, where node-specific attributes are introduced as covariates, together with the sender/receiver random effects. Other extensions of the basic model are discussed by \cite{hoff03} and \cite{hoffbili}, that bring together sender/receiver effects and latent space representations for a single network. Part of this framework was later extended by \cite{krivi}, to allow for clustering of the nodes in the latent space. In the context of dynamic networks, \cite{sewell1} model the overall sender/receiver effect in the networks, investigating whether activity (sending links) or popularity (receiving links) is more important when the edge formation process is considered.\\
Starting with latent space models, we develop a latent space approach based on the Euclidean distance to model transitivity and heterogeneity in multidimensional networks. Our aim is to extend the model by \cite{mio} to account for degree heterogeneity in multidimensional networks. Indeed, multiplex data are complex in two directions: the number of nodes and the number of views. A model that aims at describing the interactions between the actors in such a high dimensional context should explain the view-specific features, while being parsimonious with respect to the number of parameters. For this purpose, we model transitivity via a single latent space, common to the whole multiplex, assuming that the distances in such a space represent the overall association between the nodes. A single latent space is able to capture ``characterizing'' traits of the nodes, i.e. their individuality. Indeed, multiple networks represent multiple expressions of the same nodes, and the use of a single latent space helps distinguishing individual, core features from view-specific ones. Heterogeneity across different views, i.e. node-specific characteristics, will be addressed introducing node-specific sender/receiver parameters, that will account both for intra(-) and inter(-) networks degree heterogeneity.\\
The paper is organized as follows: in Section \ref{cap:cap3:sec:modello} we introduce more formally the concept of multidimensional networks and define the proposed class of models for directed multiplexes. Section \ref{cap:cap3:sec:estimation} details the estimation procedure and discusses issues of model identifiability. Section \ref{cap:cap3:sec:undirected} describes the proposed class of models for the particular case of undirected multidimensional networks. A simulation study is conducted is Section \ref{cap:cap3:sec:simulation}, to investigate the properties of the proposed class of models, and to study the performance of a novel heuristic procedure for model selection. An application to FAO trade data is presented in Section \ref{cap:cap3:sec:application}. We conclude with a discussion in Section \ref{cap:cap3:sec:disc}.

\section{The collection of latent space models for directed multiplexes}
\label{cap:cap3:sec:modello} 
A binary multidimensional network (multiplex) $\Y$ is a complex object defined by a collection of networks (also known as views). These networks can be represented by a set of $n \times n$ adjacency matrices $ \bigl\{ y^{(1)}, \dots, y^{(K)}\bigr\}=\Y $, where the index $k = 1, \dots, K$ indicates the different views. The entries in each matrix ($k^{\text{th}}$ network) can take two values, $\arch=1$ when nodes $(i,j)$ are joined by an edge, and $\arch=0$, when they are not. Views in a multiplex share the same set of nodes, whose cardinality is denoted by $n$. In the present context, nodes will be indexed by $i, j= 1, \dots, n$. 
A multiplex can be either undirected, if $\arch = y_{ji}^{(k)}$ or directed, if the different adjacency matrices are not symmetric, $\arch \neq y_{ji}^{(k)}$ for at least one $(i,j)$ couple. In general, many interesting real world multiplexes are directed, and different levels of ``symmetry'' can be observed in the adjacency matrices at hand. Notice that, even if a network is undirected, this does not imply that all the nodes have the same number of connections, that is, the same degree. Modelling the degree, or, for directed networks, the out-degree, $\sum_{j \neq i}^n y_{ij}^{(k)}$, and the in-degree, $\sum_{j \neq i}^n y_{ji}^{(k)}$, is a task that might be of interest in many empirical applications. Indeed, it can help recover the most influential, or popular, nodes in a network, or the more active ones. Further, different views may exhibit different levels of heterogeneity in the node-specific degree distribution. In this sense, multidimensional networks can be heterogeneous in two directions: within and between the views. In the present work, we introduce a class of latent space models that address transitivity and view-specific heterogeneity in the analysis of multiplex data. We start by introducing the general latent space framework for directed multidimensional networks. 
Section \ref{cap:cap3:sec:undirected} discusses their restriction to the particular case of undirected multidimensional networks.\\
In latent space models based on Euclidean distances we assume that each node is located into an unobserved $p$-dimensional Euclidean space; according to model specification, the probability of observing an edge between the dyad $(i,j)$, conditionally on the latent coordinates $\bm{z}_i, \bm{z}_j$, $i,j=1, \dots,n$, does not depend on the other nodes positions, see \cite{hoff}. We hold these assumptions in our model, together with a further one, as we assume the probability of a connection between nodes in a dyad also depends on node-specific propensity to send/receive links. In multidimensional networks, the propensity may vary with the views, as an actor could be quite popular in a network, while receiving few edges in another one. Different levels of heterogeneity in edge probabilities may depend on different levels of heterogeneity in the networks.\\
For this purpose, let $\tik$ and $\gamma_i^{(k)}$, $i=1, \dots, n$ and $k=1, \dots,K$ represent the sender and the receiver parameters for the $i^{\text{th}}$ node in the $k^{\text{th}}$ network, respectively. These parameters are introduced in the model specification to describe the propensity of a given node to send/receive edges, respectively.
The probability $p_{ij}^{(k)}$ of a connection from node $i$ to node $j$, in the $k^{\text{th}}$ network, depends on the parameters $\bigl(\theta_i^{(k)}, \gamma_j^{(k)}\bigr)$, through the following model:
\begin{equation}
 p_{ij}^{(k)} =  P\Bigl( \arch = 1 \mid \ak,\bk,\tik, \gjk ,d_{ij}\Bigr) 
 =\frac{\exp\bigl\{ f(\ak ,\tik, \gjk ) -\bk  d_{ij} \bigr\} }{1 + \exp\bigl\{ f(\ak ,\tik, \gjk)-\bk d_{ij} \bigr\} }
 \label{cap:cap3:eq:prob1}
\end{equation}
where $\bm{\alpha} = \bigl( \alpha^{(1)},\dots , \alpha^{(K)}\bigr)$ and $\bm{\beta} = \bigl(\beta^{(1)},\dots, \beta^{(K)} \bigr)$ are network-specific vector parameters and $d_{ij}$ is the squared Euclidean distance between node $i$ and node $j$ in the $p$-dimensional latent space. \\
According to the node-specific behaviours, we may define three different scenarios for each parameter:
\begin{itemize}
\item \emph{Null} (N): The nodes do not have any specific propensity to send and/or receive links; the edge probabilities reduce to a function of the distances in the latent space alone. This scenario is parametrized as $\theta_i^{(k)} = 0$ or $\gamma_i^{(k)} = 0$, $\forall$ $i,k$;
\item \emph{Constant} (C): The nodes exhibit a different propensity in sending/receiving links, but such propensities are constant over the considered views. This second scenario is parametrized by fixing $\theta_i^{(k)} = \theta_i $ or  $\gamma_i^{(k)} = \gamma_i $, $\forall$ $i,k$;
\item \emph{Variable} (V): The propensity of nodes to send and/or receive links varies across the views analysed. This last scenario may be parametrized considering $\theta_i^{(k)}$ or  $\gamma_i^{(k)}$, $\forall$ $i,k$.
\end{itemize}
Note that we assume the same type of effect (null, constant or variable) for all the views in the multiplex. While this assumption may seem a stringent one, we may observe that, in practice, assuming $\theta_i^{(k)}$ and $\gamma_i^{(k)}$ are variable we may have some nodes with null effects, others with constant effects and the remaining with variable effects. We further discuss this assumption in Section \ref{cap:cap3:sec:disc}. Table \ref{cap:cap3:table:modelli} presents a schematic taxonomy of the $9$ different models, arising from the different assumptions upon the sender and receiver effects. \begin{center}
\begin{tabular}{llllll}
\hline
   & & \multicolumn{3}{c}{ Receiver effect}\\
   & & \quad $0$ & $\gamma_{j}$ & $\gamma_{j}^{(k)}$ \\       
%   & & N & C & V \\    
   \hline
\multirow{3}{*}{Sender effect }& $0$ &  \vline  \quad NN & NC & NV \\
& $\theta_{i}$  & \vline  \quad CN & CC & CV \\
& $\theta_{i}^{(k)}$  &  \vline \quad VN & VC & VV \\
\hline
\end{tabular}
\captionof{table}{The class of models defined by the different assumptions on the sender/receiver effects.}
\label{cap:cap3:table:modelli}
\end{center}  
The impact of the sender/receiver effects on the edge probabilities can be made explicit by defining a collection of network-specific matrices $\Phi = \bigl(  \Phi^{(1)}, \dots, \Phi^{(K)}\bigr)$, with generic element defined by 
\begin{equation}
\label{cap:cap3:eq:phi}
\Bigl[ \phi_{ij}^{(k)}  \Bigr]= \Biggl[ g \bigl(   \theta_{i}^{(k)} ,   \gamma_{j}^{(k)} \bigr) \Biggr], \quad \text{with }
g(\cdot, \cdot) = 
\begin{cases} 
  1 \quad \text{if both effects are absent},  \\ 
     \theta_{i}^{(k)} \quad \text{if only the sender effect is present} ,\\ 
 \gamma_{j}^{(k)} \quad \text{if only the receiver effect is present} ,\\ 
 \frac{\theta_{i}^{(k)} +   \gamma_{j}^{(k)}}{2} \quad \text{if both effects are present}. \\  
\end{cases}
\end{equation}
We can define the function $f(\cdot)$ as:
\begin{equation}
\label{cap:cap3:eq:defIntSR}
 f( \ak ,\theta_{i}^{(k)},  \gamma_{j}^{(k)}  )  
= f \Bigl(\ak \phi_{ij}^{(k)} \Bigr) =    \alpha^{(k)}g \bigl(   \theta_{i}^{(k)} ,   \gamma_{j}^{(k)} \bigr)
\end{equation}
Equations (\ref{cap:cap3:eq:prob1}-\ref{cap:cap3:eq:defIntSR}) explicit the basic modelling assumption; within each view, the sender and the receiver effects jointly impact the view-specific intercept. Furthermore, if we assume that 
$$
\gamma_j^{(k)}, \theta_i^{(k)} \thicksim Unif(-1,1), \quad i,j=1, \dots,n
$$
we have that, differently from the additive sender and receiver effect specification, see \cite{krivi}, and as in standard latent space models, the intercepts $\ak$ still correspond, on the logit scale, to the maximum value that edge probabilities in the networks may achieve. 
Thus, recalling the definition of $\phi_{ij}^{(k)}$ in equation (\ref{cap:cap3:eq:phi}), we may notice that the sender and receiver effects can be considered as relative quantities. Indeed, inactive (or unpopular) nodes will have an effect value close to $-1$, while active (or popular) nodes will have an effect value close to $1$. Obviously, the combined effect $\phi_{ij}^{(k)}$ varies in the same range. Bounding these parameters in the interval $[-1,1]$ allows to easily interpret the differences in levels of activity (or popularity) across nodes. We must also notice that if we allow $\phi_{ij}^{(k)}$ to be negative and we do consider negative intercepts $\ak$, a fundamental problem arises. In fact, if we take $\ak<0$ and two dyads with a common node, say $(i,j)$ and $(i,l)$ with $\phi_{ij}^{(k)} < \phi_{il}^{(k)} <0$, we obtain $p_{ij}^{(k)} >p_{il}^{(k)} $ and this would violate the assumption that sender and receiver effects are directly proportional to edge probabilities. 
For this reason, we bound the intercept to be non-negative, $\ak \geq 0=LB(\alpha)$, $k =1, \dots,K$. This constraint does not alter the interpretation of the intercept and other model parameters. Indeed, fixing a lower bound for the intercept does not imply a lower bound for the edge probabilities, as the impact of the distances in the latent space may decrease this value. \\
From equation (\ref{cap:cap3:eq:defIntSR}), we may also notice that when no sender/receiver effects are present (scenario NN in Table \ref{cap:cap3:table:modelli}), the edge probability in (\ref{cap:cap3:eq:prob1}) reduces to the model specification by \cite{mio}, for which inference procedures have already been provided. In the next section, we will focus on those scenarios that include at least one effect. For these models, the edge probability formula presented in equation (\ref{cap:cap3:eq:prob1}) can be rewritten as:
\begin{equation}
\label{cap:cap3:eq:prob2}
p_{ij}^{(k)} = \frac{\exp\bigl\{ \ak \phi_{ij}^{(k)} -\bk  d_{ij} \bigr\} }{1 + \exp\bigl\{ \ak \phi_{ij}^{(k)}-\bk d_{ij} \bigr\} }
\end{equation}

\section{Estimation}
\label{cap:cap3:sec:estimation}
We propose a hierarchical Bayesian approach to parameter estimation for the latent space model proposed in Section \ref{cap:cap3:sec:modello}. The (log-)likelihood can be derived from equation (\ref{cap:cap3:eq:prob2}), 
\begin{equation}
 \ell\bigl( \bm{\alpha}, \bm{\beta}, \Phi,D \mid \Y \bigr) =\sum_{k = 1}^K \sum_{\substack{i = 1\\ j \neq i }} \ell_{ij}^{(k)}
    =\sum_{k = 1}^K  \sum_{\substack{i = 1\\ j \neq i }} \arch \log \bigl(p_{ij}^{(k)} \bigr)+ ( 1 - \arch) \log \bigl(1 -p_{ij}^{(k)} \bigr) .
  \label{cap:cap3:eq:liekli}
\end{equation}
The prior distributions for model parameters can be specified as follows:
\[
\beta^{(k)} \thicksim N_{(0, \infty)} \Bigl( \mu_{\beta}, \sigma_{\beta}^2 \Bigr), \quad \alpha^{(k)} \thicksim N_{(0, \infty)} \Bigl( \mu_{\alpha}, \sigma_{\alpha}^2 \Bigr),\quad   \bm{z}_i \thicksim MVN_p \Bigl( 0, I \Bigr), \quad \gamma_j^{(k)}, \theta_i^{(k)} \thicksim Unif(-1,1).
\]
Since $\mu_{\beta}, \mu_{\alpha}, \sigma_{\beta}^2 , \sigma_{\alpha}^2  $ are nuisance parameters whose specification could be relevant, we introduce an extra layer of dependence using the following (hyper) prior distributions:
\[
 \mu_r \mid \sigma_r^2 \thicksim N_{(0, \infty)} \Bigl( m_r, \tau_r \sigma_{r}^2 \Bigr), \quad \sigma_r^2 \thicksim \text{Inv}\chi_{\nu_r}^2,
\]
with $r = (\alpha, \beta)$. The hyperparameters $m_{\alpha}, m_{\beta}, \tau_{\alpha}, \tau_{\beta},  \nu_{\alpha}, \nu_{\beta}$ have to be specified by the user. The constraint $\bk \geq 0$, $k = 1, \dots, K$ is imposed according to the assumption that edge probabilities are inversely proportional to the distance between nodes in the latent space. A schematic representation of the hierarchical structure of the proposed model is displayed below.
\begin{figure}
\begin{center}
\begin{tabular}{cc}
      % model_pca.tex
%
% Copyright (C) 2012 Jaakko Luttinen
%
% This file may be distributed and/or modified
%
% 1. under the LaTeX Project Public License and/or
% 2. under the GNU General Public License.
%
% See the files LICENSE_LPPL and LICENSE_GPL for more details.

% PCA model

%\beginpgfgraphicnamed{model-pca}
\begin{tikzpicture}

  % Define nodes
  \node[obs]                               (y) {$y$};
  \node[latent, above=of y, xshift=-1.5cm] (a) {$\alpha$};
  \node[latent, above=of y, xshift=1.5cm]  (b) {$\beta$};
  \node[latent, right=2cm of y]            (z) {$\mathbf{z}$};
  \node[latent, left=2cm of y]  (t) {$\theta$};
  \node[latent, left=2cm of y, yshift=.9cm]  (g) {$\gamma$};
  \node[latent, above=of a, xshift=-.9cm]  (ma) {$\mu_{\alpha}$};
  \node[latent, above=of a, xshift=.9cm]  (sa) {$\sigma_{\alpha}^2$}; 
  \node[latent, above=of b, xshift=-.9cm]  (mb) {$\mu_{\beta}$};
  \node[latent, above=of b, xshift=.9cm]  (sb) {$\sigma_{\beta}^2$}; 

  % Connect the nodes
  \edge {b,a,z,g,t} {y} ; %
  \edge{ma,sa} {a};
  \edge{sa}{ma};
  \edge{mb,sb}{b};
  \edge{sb}{mb};

  % Plates
 % \plate {yx} {(x)(y)} {$N$} ;
  %\plate {} {(w)(y)(yx.north west)(yx.south west)} {$M$} ;

\end{tikzpicture}
%\endpgfgraphicnamed

%%% Local Variables: 
%%% mode: tex-pdf
%%% TeX-master: "example"
%%% End:  
\end{tabular}
\end{center}
\caption{Hierarchy structure for the models.}
\end{figure}
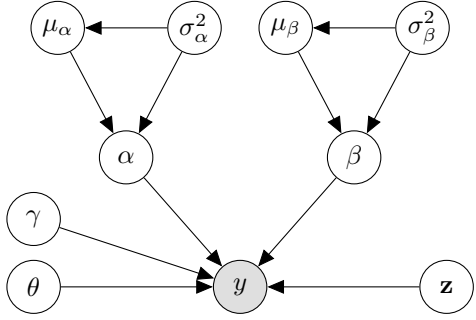

\subsection{Identifiability}
\label{cap:cap3:sec:identif}
As discussed in \cite{mio}, to ensure parameter identifiability for the basic latent space model, one out of the $K$ parameters $\ak$ and $\bk$ must be fixed. The corresponding network is then referred to as the ``reference'' network. In the present context, fixing these two parameters is not enough, as the multiplicative effect of $\phi_{ij}^{(k)}$ may still cause problems. To avoid such issues, if both effects are present, one sender and one receiver effect should be fixed in each view, and the corresponding nodes will be considered as ``reference'' nodes. More in details, when the effect is variable, we propose to choose as reference, in each network, the $i^{\text{th}}$ node with the highest observed out-degree (or in-degree) and fix  $\theta_i^{(k)}= 1$ (respectively $\gamma_i^{(k)}= 1$).
Instead, when the effect is constant, we propose to select the $i^{\text{th}}$ node with highest observed mean out-degree (or mean in-degree) and fix $\theta_i= 1$ (respectively $\gamma_i= 1$).
Fixing the sender/receiver parameter for one node to $1$ does not change the interpretation of the model, as we are most interested in ordering the nodes with respect to the sender/receiver effects rather than in deriving precise point estimates.

\subsection{MCMC algorithm}
\label{cap:cap3:sec:MCMC}
We propose an MCMC algorithm to estimate parameters for the latent space model proposed in section \ref{cap:cap3:sec:modello}. The algorithm iterates over the updated estimates of model parameters, latent coordinates and nuisance parameters; since full conditionals are available in closed form only for the latter ones, we use a Metropolis-Hastings step to update the other estimates.
Full conditional distributions for the nuisance parameters, proposal distributions for the network-specific parameters and the latent coordinates are described in Appendix \ref{cap:appendice2:app2}, together with proposal distributions for the sender/receiver effects. 
The adopted procedure starts by simulating a new value for each nuisance parameter; then, it proposes a new value for the network-specific parameters $\ak$ and $\bk$, with a joint MH step on each network. A further MH step sequentially proposes new latent coordinate values. As the likelihood in equation (\ref{cap:cap3:eq:liekli}) is invariant to rotations and translations of the latent coordinates, when a new set of positions is defined, Procrustes transformation is employed to check whether this new set is just a simple transformation of the previous solution. If so, the proposed set is discarded in favour of the previous one.
After that, sender/receiver parameters are updated sequentially on the nodes, but jointly over the different networks via an additional MH step. The joint update is performed to speed up the calculations, given the high number of model parameters. When all the effects are updated, a new solution of $\Phi^{(k)}$ is available. The proposed algorithm scales quadratically with the number of nodes $n$ and linearly with the number of networks.\\
Latent coordinates are initialized via multidimensional scaling on the average geodesic distances calculated over the different networks. Squared Euclidean distances are then computed on the starting latent positions, and are used to perform a logistic regression for the adjacency matrix binary entries to get starting values  for the intercept and the coefficient parameters $\ak$ and $\bk$: Sender and receiver parameters are initialized in a non-informative way, by fixing them to $0$.\\
Edge specific covariates, either constant or variable across the networks can be easily incorporated in the proposed model for the edge probabilities. Also, the model can be easily extended to deal with the presence of missing edges/nodes in the data. The full conditional and proposal distributions for model parameters presented in Appendix \ref{cap:appendice2:app2} refer to the most general case of missing data and edge-specific covariates. 

\section{Undirected network}
\label{cap:cap3:sec:undirected}
In the particular case of undirected networks, out-degrees and in-degrees are identical for each node, $\sum_{i} y_{ij}^{(k)} = \sum_{i} y_{ji}^{(k)}$ $\forall i = 1, \dots, n$. The framework proposed in section \ref{cap:cap3:sec:modello} can be easily modified to deal with undirected multidimensional networks by imposing the constrain: $\theta_i^{(k)} = \gamma_i^{(k)}=\delta_i^{(k)}$. According to such assumption, the edge probability equation (\ref{cap:cap3:eq:prob2}) can be rewritten as:
\begin{equation}
 P\Bigl( \arch = 1 \mid  \ak,\bk,\delta_i^{(k)}, \delta_j^{(k)},d_{ij}\Bigr)
= \frac{\exp\bigl\{ \ak g(\delta_i^{(k)} ,\delta_j^{(k)}) -\bk  d_{ij} \bigr\} }{1 + \exp\bigl\{ \ak g(\delta_i^{(k)} ,\delta_j^{(k)})-\bk d_{ij} \bigr\} }
\label{cap:cap3:eq:socprob}
\end{equation}
The effect $\delta_i^{(k)}$, if not null, can either be variable across the different networks, $\delta_i^{(k)}$, or constant, $\delta_i^{(k)} = \delta_i$, for $k=1, \dots, K$ and $i = 1, \dots, n$. Appendix \ref{cap:appendice2:app2} provides the reader with the proposal distributions used to estimate these model parameters. 

\section{Simulations}
\label{cap:cap3:sec:simulation}
We defined a simulation study to evaluate the performance of the proposed estimation procedure, where we examine the behaviour of model parameter estimates for models NC, NV, CC, CV and VV. 
For each one of the simulated multiplexes ($10$ for each model), we fit the ``true'' model, that is, the model a given multiplex was simulated from. These five models are chosen to investigate the properties of the estimators when the true scenarios refer to different numbers of parameters. A first scenario of simulated multidimensional networks, \textbf{B1}, has dimensions $(n=50,K=5)$. A second larger one, \textbf{B2}, has dimensions similar to those of  the multiplex that we analyse in the application, see section \ref{cap:cap3:sec:application}, namely $(n=50,K=10)$.\\
In both these simulation scenarios, we set $\alpha^{(1)} = 2$ and $\beta^{(1)}=1$. The prior parameters are $\nu_{\alpha}=\nu_{\beta} = 3$, $m_{\alpha}=2$, $m_{\beta}=0$, $\tau_{\alpha}=\tau_{\beta} = (K-1)/K$ and $p=2$. Small variations of these values have been also tried and did not affect the results. Also, there are no missing edges in the simulated data. The MCMC algorithm run for 60000 iterations with a burn in of 15000. \\
Table \ref{cap:cap3:table:DCvalues} shows the average values of the distance correlation computed between the simulated and the estimated edge probabilities, for all the different networks composing the multiplexes in scenarios \textbf{B1} and \textbf{B2}. As we may observe, the correlations are always high, proving that we are able to recover the edge probabilities with good quality, regardless of the true model, the view or the multiplex size. Also, last column of Table \ref{cap:cap3:table:DCvalues} reports the average values of the Procrustes transformation between the simulated and the estimated latent space coordinates. As for the edge probabilities, we notice that the latent coordinates are appropriately recovered. To evaluate the estimates of the sender/receiver parameters in the different models, we compute, for each network, the Spearman correlation coefficient between the simulated and the estimated parameters; this choice is used as we are mainly interested in recovering the nodes ordering, with respect to the two estimated effects. Indeed, sender and receiver parameters vary in the relatively small interval $(-1,1)$, and the exact numerical values may not be of interest. Table \ref{cap:cap3:table:GammaCvalues} reports, for each network in the simulated multiplexes, the average values of the Spearman correlation coefficient between the simulated and the estimated receiver parameters, $\gamma_i^{(k)}$. These values are always much greater than $0.5$, with a couple of exceptions for some networks in the models with higher complexity, CV and VV. However, as we comment in Table \ref{cap:cap3:table:DCvalues}, this does not impact the edge probabilities recovering. Table \ref{cap:cap3:table:ThetaCvalues} reports the average values of the Spearman correlation coefficient between the simulated and the estimated sender parameters, $\theta_i^{(k)}$. The behaviour of these estimates complies with those of the receiver parameters previously discussed, proving that the effects, when both present, are estimated with similar quality.\\
We refer to Appendix \ref{cap:appendice2:app2} for the $\ak$ and the $\bk$ estimates. The intercepts are recovered within a $95\%$ credible interval, with two limited exceptions which occur when the simulated values are ``extreme''. However, the ordering between the different intercepts in a given multiplex is always recovered. Instead, the $\bk$ coefficient tends to be overestimated. Also in this case, the ordering of coefficients in a multiplex is correctly recovered. The overestimation of this coefficient may be caused by a corresponding underestimation of the latent distances, as the simulated and estimated products $\bk d_{ij}$ are always well recovered. Precise point estimates of all the parameters are quite hard to recover, due to the large number of parameters in the models and the ``multiplicative'' parametrization adopted. However, the aim of this class of latent space models is to describe different features of a multiplex by comparing nodes and networks. This intent is met, as we are always able to recover the corresponding orderings. 
\begin{center}
\scalebox{.8}{
\begin{tabular}{@{}lllllllllllll@{}}
\toprule
       & & $k=1$ & $k=2$ & $k=3$ & $k=4$ & $k=5$ & $k=6$ & $k=7$ & $k=8$ & $k=9$ & $k=10$ & PC \\   
    \hline   
    &     &  \vline \quad {\color{white}0.65}   &  &  &  & &  &  &  &  &  & \\
\multirow{5}{*}{B1  }&  NC   & \vline \quad 0.86 & 0.84  & 0.86 & 0.87 & 0.85 & - & - & - & - & - & 0.94\\
&  NV   &  \vline \quad 0.87 & 0.87  & 0.87 & 0.85 & 0.86 & - & - & - & - & - & 0.92\\
&  CC   & \vline \quad 0.89 & 0.89  & 0.90 & 0.91 & 0.87 & - & - & - & - & - & 0.96\\
&  CV   & \vline \quad 0.84 & 0.81  & 0.83 & 0.84 & 0.85 & - & - & - & - & - & 0.90\\
&  VV   & \vline \quad 0.83 & 0.81  & 0.80 & 0.81 & 0.81 & - & - & - & - & - & 0.93\\
&     &  \vline \quad {\color{white}0.65}   &  &  &  & &  &  &  &  &  & \\
\multirow{5}{*}{B2  }&  NC   & \vline \quad 0.89 & 0.84 & 0.83 & 0.83 & 0.81 & 0.92 & 0.84 & 0.81 & 0.81 & 0.82 & 0.85\\
&  NV   & \vline \quad 0.83 & 0.80  & 0.79 & 0.76 & 0.81 & 0.78 & 0.88 & 0.71 & 0.75 & 0.75 & 0.93\\
&  CC   & \vline \quad 0.83 & 0.87  & 0.91 & 0.84 & 0.90 & 0.91 & 0.91 & 0.84 & 0.86 & 0.91 & 0.90\\
&  CV   & \vline \quad 0.75 & 0.86 & 0.87 & 0.76 & 0.86 & 0.88 & 0.87 & 0.86 & 0.83 & 0.87 & 0.91\\
&  VV   & \vline \quad 0.73 & 0.72 & 0.82 & 0.85 & 0.88 & 0.79 & 0.79 & 0.79 & 0.86 & 0.84 & 0.90\\
\bottomrule
\end{tabular}
}
\captionof{table}{Simulation study. Distance correlation between the simulated and the estimated edge-probabilities. Last column ($PC$) shows the Procrustes correlation between the simulated and the estimated latent space coordinates. }
\label{cap:cap3:table:DCvalues}
\end{center}
\begin{center}
\scalebox{.8}{
\begin{tabular}{@{}lllllllllllll@{}}
\toprule
       & & $k=1$ & $k=2$ & $k=3$ & $k=4$ & $k=5$ & $k=6$ & $k=7$ & $k=8$ & $k=9$ & $k=10$  \\   
      \hline   
    &     &  \vline \quad {\color{white}0.65}   &  &  &  & &  &  &  &  &  & \\
\multirow{5}{*}{B1  }&  NC   & \vline \quad 0.84 & 0.84  & 0.84 & 0.84 & 0.84 & - & - & - & - & - \\
&  NV   &\vline \quad 0.85 & 0.80  & 0.86 & 0.92 & 0.91 & - & - & - & - & - \\
&  CC   &\vline \quad 0.84 & 0.84  & 0.84 & 0.84 & 0.84 & - & - & - & - & - \\
&  CV   &\vline \quad 0.80 & 0.86  & 0.81 & 0.73 & 0.6 & - & - & - & - & - \\
&  VV   &\vline \quad 0.67 & 0.61  & 0.50 & 0.64 & 0.75 & - & - & - & - & - \\
&     &  \vline \quad {\color{white}0.65}   &  &  &  & &  &  &  &  &  \\
\multirow{5}{*}{B2  }&  NC   &\vline \quad 0.94 & 0.94 & 0.94 & 0.94 & 0.94 & 0.94 & 0.94 & 0.94 & 0.94 & 0.94 \\
&  NV   &\vline \quad 0.92 & 0.86  & 0.84 & 0.74 & 0.84 & 0.83 & 0.92 & 0.61 & 0.80 & 0.79 \\
&  CC   &\vline \quad 0.84 & 0.84  & 0.84 & 0.84 & 0.84 & 0.84 & 0.84 & 0.84 & 0.84 & 0.84 \\
&  CV   &\vline \quad 0.51 & 0.66 & 0.83 & 0.50 & 0.73 & 0.82 & 0.82 & 0.85 & 0.61 & 0.79 \\
&  VV   &\vline \quad 0.61 & 0.48 & 0.81 & 0.83 & 0.86 & 0.58 & 0.54 & 0.57 & 0.87 & 0.59 \\
\bottomrule
\end{tabular}
}
\captionof{table}{Simulation study. Spearman correlation index between the simulated and the estimated receiver effects. }
 \label{cap:cap3:table:GammaCvalues}
\end{center}  
\begin{center}
\scalebox{.8}{
\begin{tabular}{@{}lllllllllllll@{}}
\toprule
       & & $k=1$ & $k=2$ & $k=3$ & $k=4$ & $k=5$ & $k=6$ & $k=7$ & $k=8$ & $k=9$ & $k=10$  \\   
       \hline
 \multirow{5}{*}{B1  }&  & \vline \quad {\color{white}0.65}    &  &  &  & &  &  &  &  &  & \\    
& CC    & \vline \quad 0.84 & 0.84  & 0.84 & 0.84 & 0.84 & - & - & - & - & - \\
&  CV    & \vline \quad 0.90 & 0.90 & 0.90 & 0.90 & 0.90 & - & - & - & - & - \\
&  VV    & \vline  \quad 0.58 & 0.57  & 0.55 & 0.58 & 0.59 & - & - & - & - & - \\
\multirow{5}{*}{B2  } &     & \vline \quad {\color{white}0.65}  &  &  &  & &  &  &  &  &  \\
&  CC   & \vline  \quad 0.78 & 0.78 & 0.78 & 0.78 & 0.78 & 0.78 & 0.78 & 0.78 & 0.78 & 0.78 \\
&  CV   & \vline \quad 0.79 & 0.79 & 0.79 & 0.79 & 0.79 & 0.79 & 0.79 & 0.79 & 0.79 & 0.79 \\
&  VV &    \vline \quad 0.65 & 0.66 & 0.66 & 0.65 & 0.66 & 0.66 & 0.66 & 0.66 & 0.67 & 0.65 \\
\bottomrule
\end{tabular}
}
\captionof{table}{Simulation study. Spearman correlation between the simulated and the estimated sender effects, by simulation scenario and true model structure.}
 \label{cap:cap3:table:ThetaCvalues}
\end{center}

\subsection{An heuristic procedure for model selection}
\label{cap:cap3:sec:euristico}
In Section \ref{cap:cap3:sec:modello} we have proposed a class of models for multidimensional networks, with 9 different specifications of the sender/receiver effects. In general, the issue of model selection can be addressed in two different ways. A first approach is that of an expert who, based on some prior knowledge on the data, suggests which particular model should be used. A second, more common, approach, is that of choosing the ``best'' model using some selection criteria. 
In the present context, the estimation of the nine models, on a specific observed multiplex may request some (computational) time, especially when the number of nodes is large. 
Hence, it could be convenient to have some idea on which model to estimate on a priori basis. In particular, we propose to determine the model to fit on the basis of some summary statistics calculated on the current multidimensional network data. The idea of using summary statistics to aid the model selection was introduced by \cite{GOFnet}. In this work, the authors proposed a graphical goodness of fit procedure to compare structural statistics of the observed network with the corresponding statistics on networks simulated from a given model. Although this procedure allows to compare different models without the need to fit them, it relies on graphical comparisons, which may be subjective or difficult when the number of models to compare is large. A more automated model selection procedure is that discussed by \cite{Pudlo:2015} in the context of ABC algorithms. Here, the model selection problem is rephrased as a classification one, with random forests used to predict, in advance of fitting, which model would be the most appropriate for the data.\\
We propose to summarize observed multiplex data using in-degrees and out-degrees correlations. The general idea is that of computing the correlations among out-degrees/in-degrees. This could serve as a proxy of the heterogeneity within and between the views. 
Let us denote by $\mathbf{S} = [s_{ik}]$ the matrix of the observed out-degrees and by $\mathbf{R} = [r_{ik}]$ the matrix of the observed in-degrees; both matrices have dimension $n \times K$, where $n$ denotes the number of nodes and $K$ the number of networks. Then, the matrices $c_{s_k}$ and $c_{r_k}$, of dimension $K \times K$, contain the values of the correlation between the sender/receiver effects across views. Instead, the matrices $c_{s_i}$ and $c_{r_i}$, of dimension $n \times n$, include the correlations between the sender/receiver effects across nodes. Let us now define $\bar{c}_{s_k}$, $\bar{c}_{r_k}$, $\bar{c}_{s_i}$ and $\bar{c}_{r_i}$ the mean values among all the cells of the matrices introduced above and $sd(c_{s_k})$, $sd(c_{r_k})$, $sd(c_{s_i})$ and $sd(c_{r_i})$ the corresponding standard deviations.
We use such quantities to choose which type of model has to be estimated, among those proposed in Section \ref{cap:cap3:sec:modello}.
The idea is that observed multiplexes with similar values of $\bar{c}_{s_k}$ and $\bar{c}_{r_k}$, and of $\bar{c}_{s_i}$ and $\bar{c}_{r_i}$, could have similar types of sender and receiver effects. On the contrary, a multiplex that exhibit conflicting values of sender/receiver correlation among views or among nodes might come from a model where the two effects are different. In the latter case, the higher the discrepancy between sender and receiver ``node'' correlations, the higher the chance that the underlying model has two most different types of effect, that is, null and variable. 
The variability between nodes correlations (for which the standard deviations summary statistics serve as proxies) may be used to discriminate between different model complexities, that is null, constant or variable node-specific effects. Formally, the proposed procedure works as described below.\\
Given an observed multidimensional network with $n$ nodes and $K$ views, we propose to simulate $T$ multiplexes from equation \ref{cap:cap3:eq:prob2}, for each one of the nine models. For each simulated dataset, the correlation summary statistics ($\bar{c}_{s_k},\bar{c}_{r_k},\bar{c}_{s_i},\bar{c}_{r_i},sd(c_{s_k}), sd(c_{r_k}), sd(c_{s_i}),sd(c_{r_i})$) are computed. Such summary statistics are later employed to train a Linear Discriminant Analysis (LDA) classifier, using \emph{mclust} \citep{mclust}.
The proposed procedure has been tested for multidimensional networks with $n = (50, 75, 100)$ nodes and $K=(3,5,10)$ networks, with $T=5000$. Table \ref{cap:cap3:table:cvHeu} reports the cross validation error for the classifier, for different $n$ and $K$ values. The classification errors are quite low in all scenarios, and they decrease with increasing $K$. Indeed, having multiple replicates of the networks may help distinguishing the type of network/node-specific effects. To further test the proposed procedure, we have used the estimated classifier to predict the model-type of other $5000$ multidimensional networks, simulated independently from those used to train the classifier. Table \ref{cap:cap3:table:cvAccu} reports the average accuracy of the classifier on the test data over different $(n, K)$ values, for each one of the models. The accuracy values on the diagonal show that the proposed procedure has a larger discriminative power among the competing models.
Also, when it fails to select the right model, it proposes a model ``near'' to the ``true'' one. For example, when multiplexes are simulated from a $VV$ model, the procedure recovers the true one $96\%$ of the times, the $CC$ model $2\%$ and the $CV$-$VC$ models $2\%$ of the times. \\
Using an heuristic procedure to choose among the models can help reducing drastically the computing effort. Moreover, the proposed framework allows the implementation of dimension-specific classifier, as the scheme and the corresponding classification depend directly on the dimension of the observed multidimensional network, $n,K$. 
\begin{center}
\begin{tabular}{@{}ccccccccc@{}}
\toprule
\multicolumn{3}{c}{$n=50$} & \multicolumn{3}{c}{$n=75$} &\multicolumn{3}{c}{$n=100$}\\ $K=3$ & $K=5$ & $K=10$ & $K=3$ & $K=5$ & $K=10$ & $K=3$ & $K=5$ & $K=10$ \\
\midrule
0.152 & 0.059 & 0.024 & 0.076 & 0.030 & 0.016 & 0.051 & 0.023 & 0.013 \\
\bottomrule
\end{tabular}
\captionof{table}{Cross validation error for the classifier, for different $n$ and $K$ values.}
\label{cap:cap3:table:cvHeu}
\end{center}  

\begin{center}
\scalebox{0.8}{
\begin{tabular}{@{}rrccccccccc@{}}
\toprule
& & \multicolumn{9}{c}{Predicted} \\
& & $NN$ & $CN$ & $NC$ & $CC$ & $VN$ & $NV$ & $VC$ & $CV$ & $VV$\\
\midrule
\multirow{9}{*}{Class} & $NN$ \quad \vline & \textbf{93} (2) $\%$ &  1 (0) $\%$ & 0 (0) $\%$ & 6 (1) $\%$ & 0 (0) $\%$ & 0 (0) $\%$ & 0 (0) $\%$ & 0 (0) $\%$ & 0 (0) $\%$ \\
 & $CN$ \quad \vline & 0 (0) $\%$ & \textbf{97} (4) $\%$ & 0 (0) $\%$ & 2 (2) $\%$ & 0 (0) $\%$ & 0 (0) $\%$ & 0 (0) $\%$ & 1 (1) $\%$ &  0 (0) $\%$   \\
 & $NC$ \quad \vline & 0 (0) $\%$ & 0 (0) $\%$ & \textbf{97} (3) $\%$ & 2 (2) $\%$ & 0 (0) $\%$ & 0 (0) $\%$ & 1 (1) $\%$ & 0 (0) $\%$ & 0 (0) $\%$\\
 & $CC$ \quad \vline & 6 (2) $\%$ & 1 (1) $\%$ & 1 (1) $\%$ & \textbf{88} (7) $\%$ & 0 (0) $\%$ & 0 (0) $\%$ & 1 (1) $\%$ & 1 (2) $\%$ & 2 (1) $\%$ \\
 & $VN$ \quad \vline & 0 (0) $\%$ & 0 (0) $\%$ & 0 (0) $\%$ & 0 (0) $\%$ & \textbf{97} (3) $\%$ & 0 (0) $\%$ & 3 (2) $\%$ & 0 (0) $\%$ & 0 (0) $\%$\\
 & $NV$ \quad \vline & 0 (0) $\%$ & 0 (0) $\%$ & 0 (0) $\%$ & 0 (0) $\%$ & 0 (0) $\%$ & \textbf{97} (3) $\%$ & 0 (0) $\%$ & 3 (2) $\%$ & 0 (1) $\%$\\
 & $VC$ \quad \vline & 0 (0) $\%$ & 0 (0) $\%$ & 0 (1) $\%$ & 1 (2) $\%$ & 3 (2) $\%$ & 0 (0) $\%$ & \textbf{95} (6) $\%$ & 0 (0) $\%$ & 1 (2) $\%$\\ 
 & $CV$ \quad \vline & 0 (0) $\%$ & 0 (1) $\%$ & 0 (0) $\%$ & 1 (2) $\%$ & 0 (0) $\%$ & 3 (2) $\%$ & 0 (0) $\%$ & \textbf{95} (6) $\%$ & 1 (2) $\%$\\ 
 & $VV$ \quad \vline & 0 (0) $\%$ & 0 (0) $\%$ & 0 (0) $\%$ & 2 (1) $\%$ & 0 (0) $\%$ & 0 (0) $\%$ & 1 (2) $\%$ & 1 (2) $\%$ & \textbf{96} (5) $\%$ \\
\bottomrule
\end{tabular}
}
\captionof{table}{Average accuracy of the classifier on the test data over different $(n, K)$ values, for each one of the nine models. Standard deviations are reported in brackets. }
\label{cap:cap3:table:cvAccu}
\end{center}

\section{FAO trade data}
\label{cap:cap3:sec:application}
\begin{figure}[h]%
    \centering
    \subfloat[Estimated sender parameters and observed mean out-degrees. The grey lines represent out-degree standard deviations.]{{\includegraphics[scale = .42]{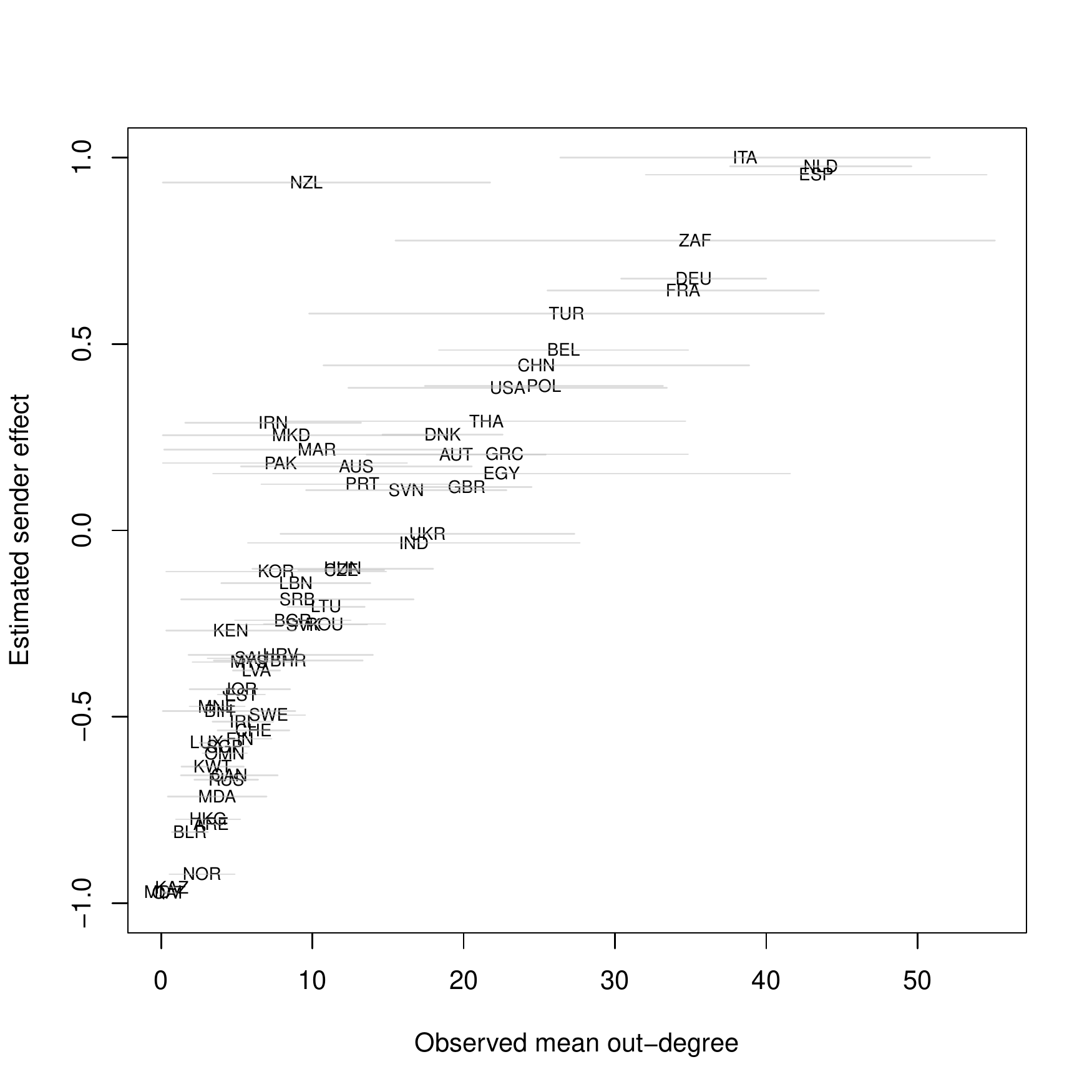} }}
   \subfloat[Estimated latent space. The segments link New Zeland to the countries it exports to, with different colors corresponing to different fruit markets.]{{\includegraphics[scale=0.42]{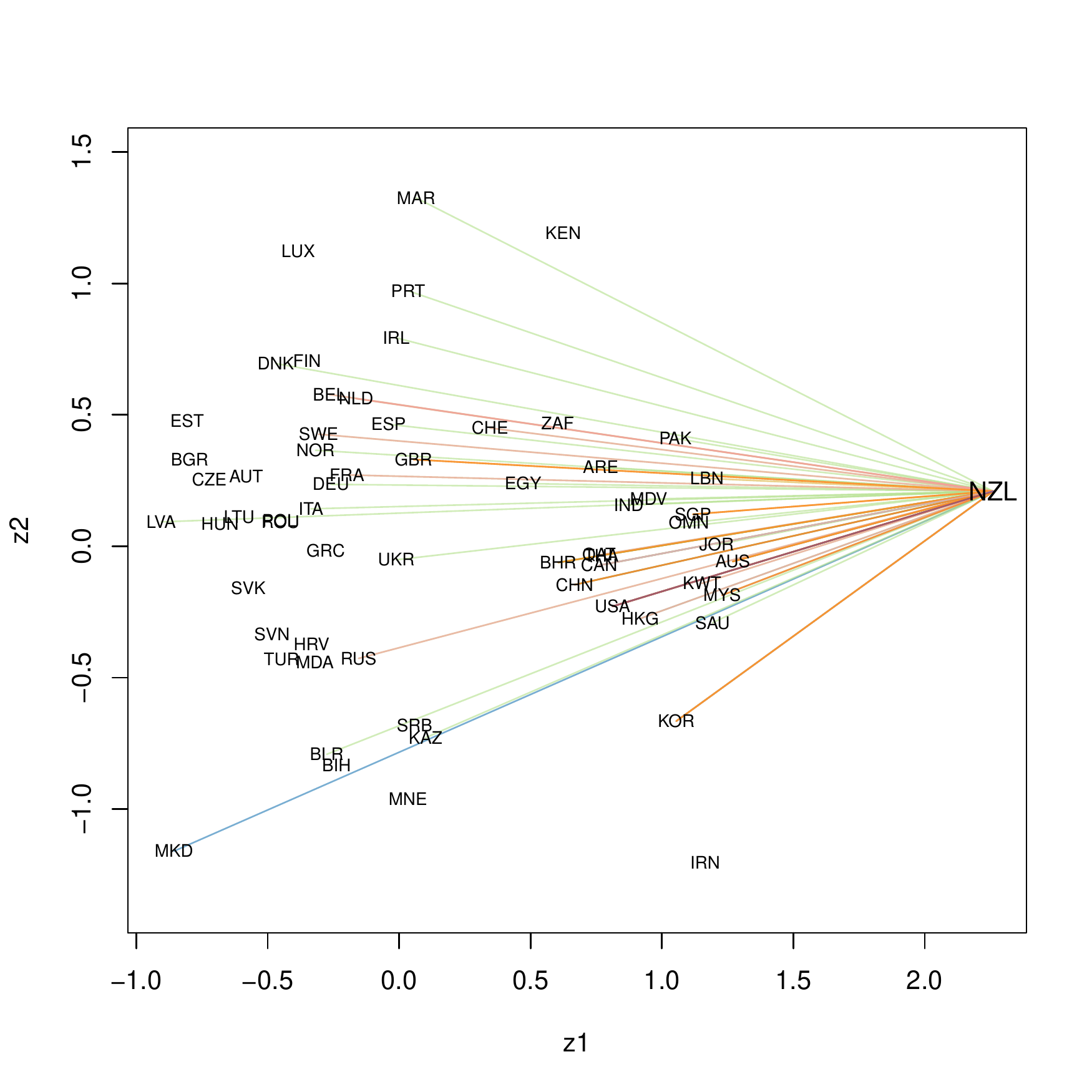} }}%
    \caption{Fruit multiplex. Estimated sender parameters and observed mean out-degrees, and latent space representation of New Zeland exports.}%
    \label{cap:cap3:fig:latPosFruttaNZ}%
\end{figure}

\begin{figure}[h]%
    \centering
    \subfloat[Estimated latent coordinates. The segment represented at the bottom of the plot displays the median value of the estimated distances distribution.]{{\includegraphics[scale = .42]{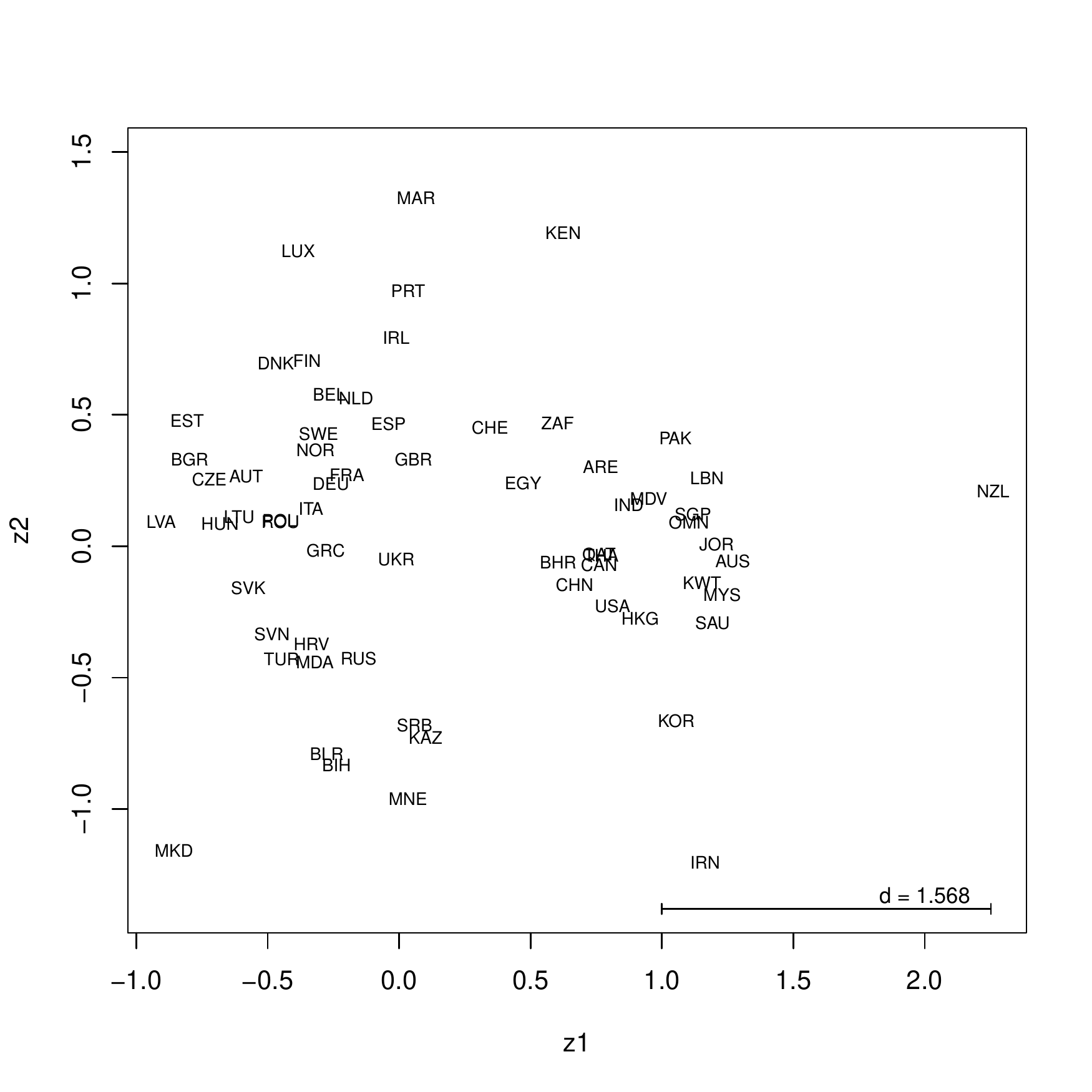} }}
   \subfloat[Geographical coordinates. Estimated latent coordinates are reported in grey in the background, for comparison.]{{\includegraphics[scale=0.42]{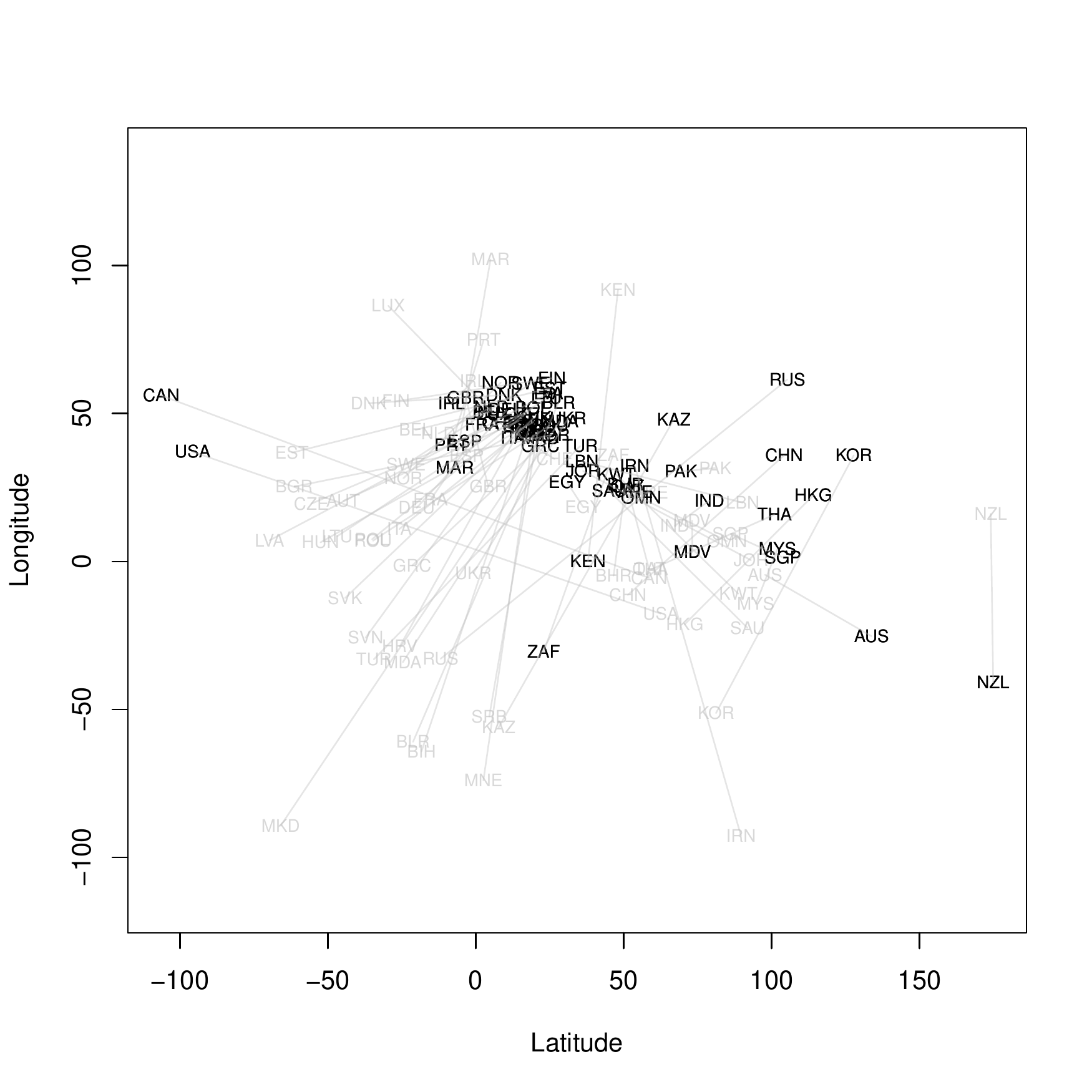} }}%
    \caption{Fruit multiplex. Estimated latent coordinates for the countries and geographical coordinates.}%
    \label{cap:cap3:fig:latPosFrutta}%
\end{figure}

\begin{figure}[h]%
    \centering
    \subfloat[First quartile, $d(i,j) = 0.385$.]{{\includegraphics[scale = .26]{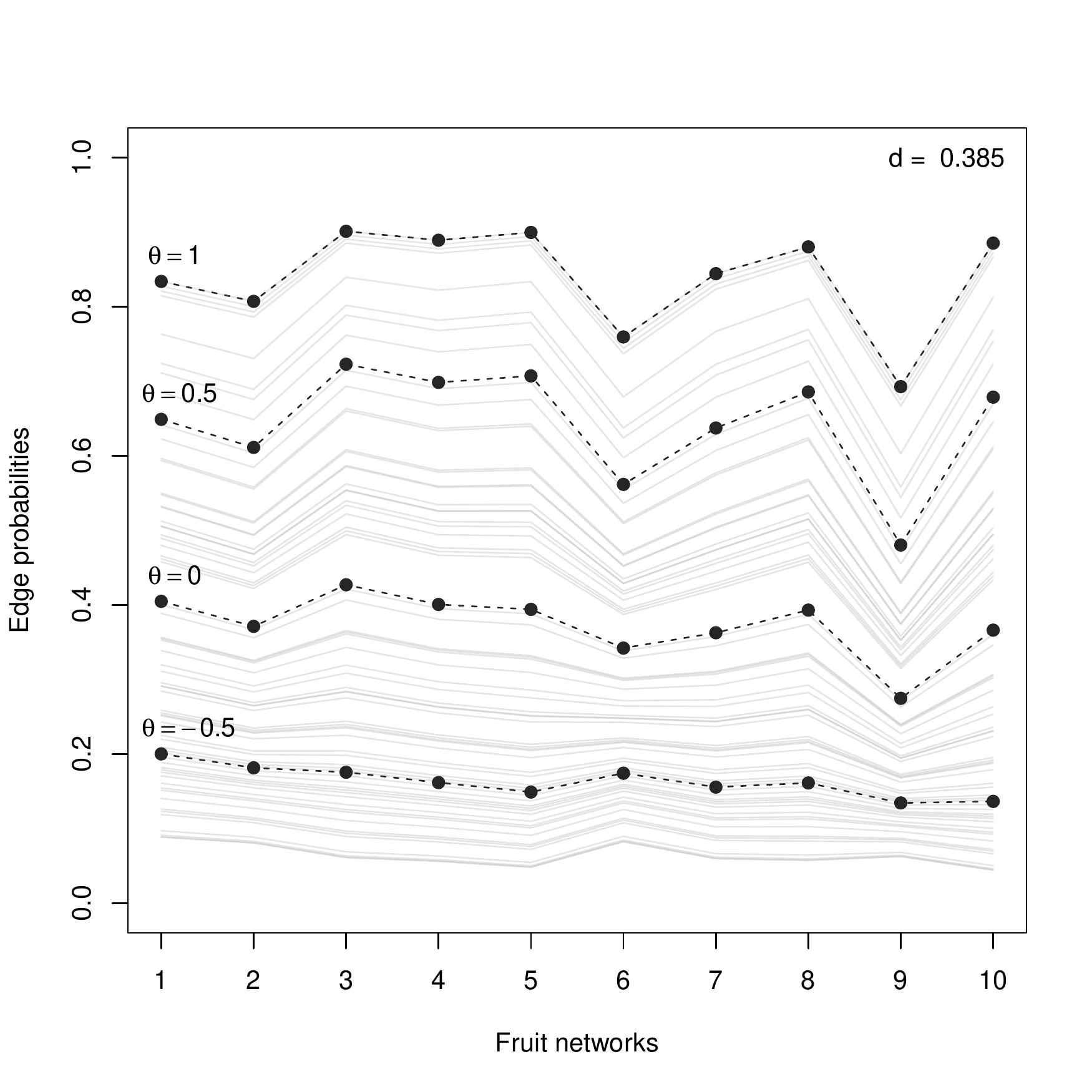} }}
   \subfloat[Median, $d(i,j) = 1.274$.]{{\includegraphics[scale=0.26]{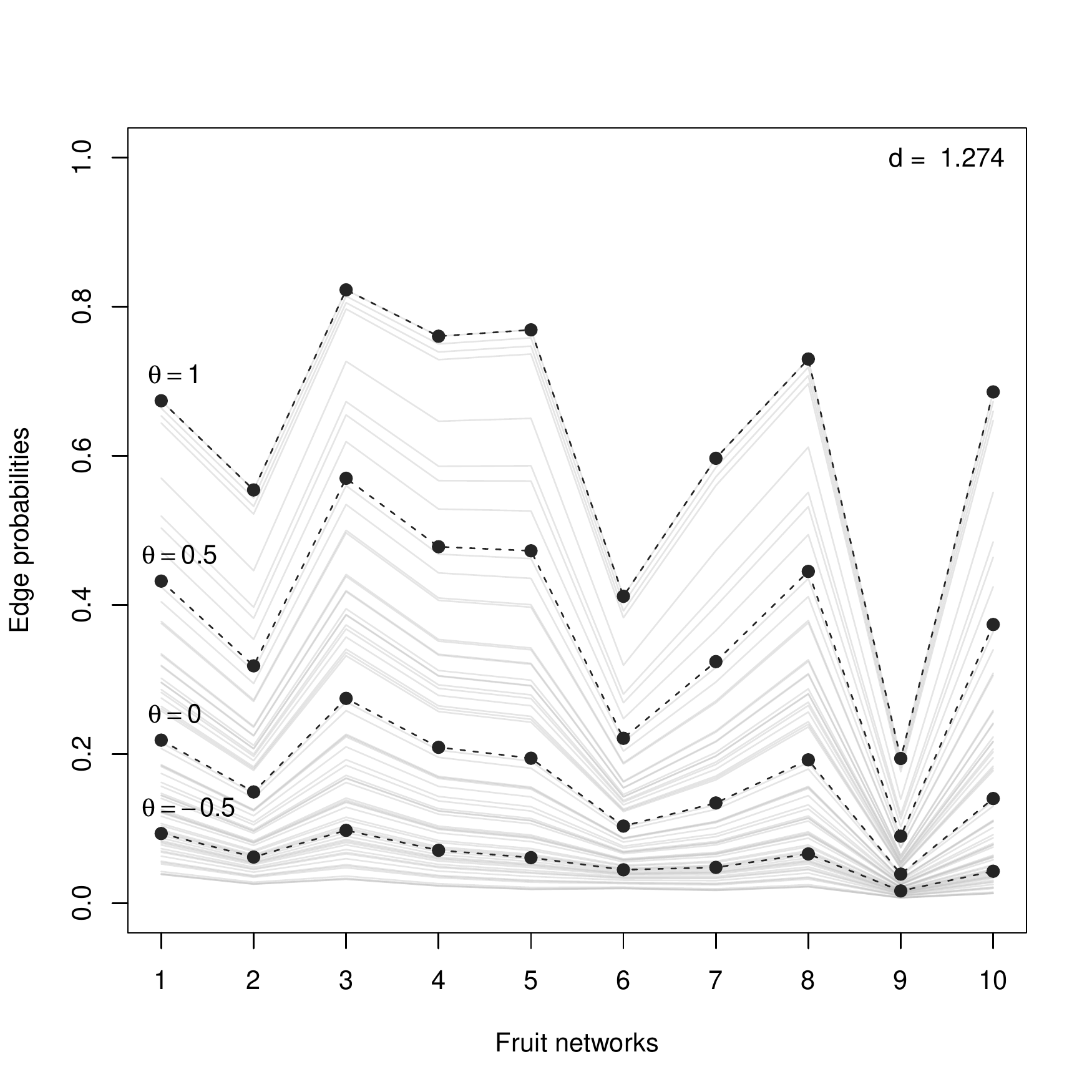} }}%
    \subfloat[Third quartile, $d(i,j) = 2.281$.]{{\includegraphics[scale=0.26]{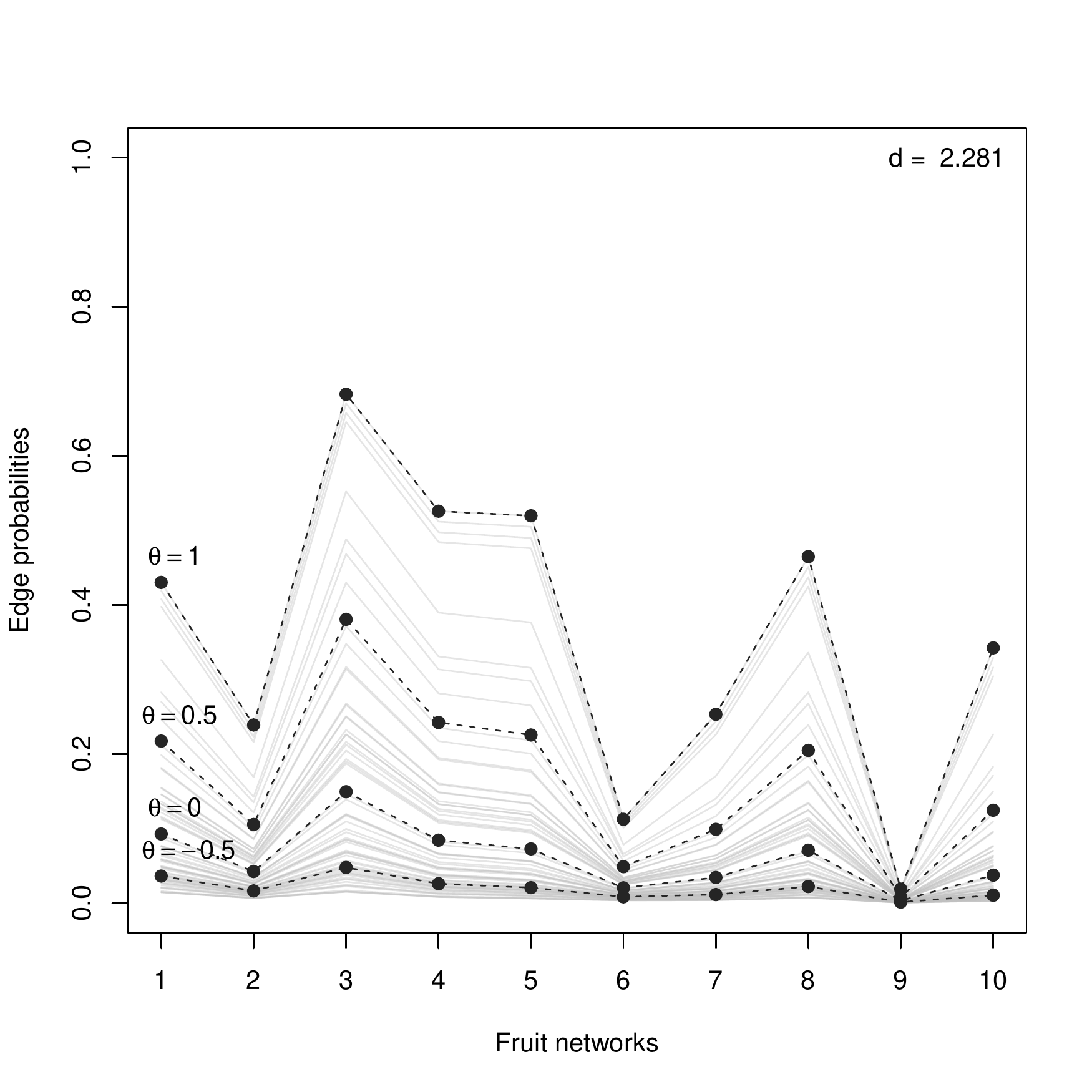} }}%
    \caption{Fruit networks. Estimated probabilities in the multiplex, for different values of the estimated distances (first quartile, median and third quartile) and for $\theta_{i} = (-0.5, 0, 0.5, 1)$. The grey lines in the background represent the country-specific edge probabilities, obtained considering the estimated sender parameters.}%
    \label{cap:cap3:fig:probFrutta}%
\end{figure}

The application deals with FAO food and agricultural trade data, measuring annual import/exports between countries. The data are available at FAO website \citep{FAOSTAT}, and the most recent subset refers to 2013. Here we consider the fruit sub-market, in particular, fresh fruit, as fresh items are the most traded internationally. For illustrative reasons, we consider a restricted number of fruits, by choosing 10 out of the most commonly consumed and traded goods: ``Grapes'', ``Watermelons'', ``Apples'', ``Oranges'', ``Pears'', ``Bananas'', ``Pineapples'', ``Tangerines, mandarins, clementines, satsuma'', ``Plantains'' and ``Grapefruit (inc. pomelos)''. The original data register the volume of the trade, that is, the quantity traded among each couple of countries; however, we focus on the presence/absence of an import/export relation between couples of countries. Our aim is to verify whether close countries are more likely to trade, by comparing the estimated latent coordinates of the countries with the geographical ones. Also, we can address which countries are the most relevant in the exchange of fruits, via the estimated node-specific effect parameters, and whether their relevance is constant throughout the different markets. The original number of trading countries in the data is large, more than $200$, but not all of them trade in all of the markets. Thus, to avoid the presence of isolated node-countries (countries with no links) and to guarantee an easy and feasible representation of the results, we focus on a subgroup of 64 countries, reported in Table \ref{cap:appendice2:tab:nomiPaesiapp}, \ref{cap:appendice2:app1}. To define such a sub-sample, we have considered the median number of countries with which a country trades (equal to $7$) and removed all the countries with a value under the median. We end up considering a multidimensional networks with $n=64$ nodes and $K=10$ networks. \\
The observed densities range from $0.10$ (Plantains market) to $0.28$ (Apples market), with a mean of $0.20$. Also, the associations\footnote{
The association between any two adjacency matrices, $k,l=1, \dots, K$, is computed comparing the total number of concordant cells between the two matrices and the total number of cells: \[As \bigl( \mathbf{Y}^{(k)} ,\mathbf{Y}^{(k)}  \bigr) = \frac{ \sum_{i,j}^n \mathbf{I} \bigl( y_{ij}^{(k)} =  y_{ij}^{(l)}\bigr)  }{ \sum_{i,j}^n \mathbf{I} \bigl( y_{ij}^{(k)} =  y_{ij}^{(l)}\bigr) + \sum_{i,j}^n \mathbf{I} \bigl( y_{ij}^{(k)} \neq y_{ij}^{(l)}\bigr)}.
\]} between couples of adjacency matrices are quite high, ranging in between $0.8$ and $0.9$, suggesting that countries tend to import/export fruits from/to a relatively constant set of partners. As the observed out-degrees and in-degrees present a strong association (see the Supplementary materials), the data are a good candidate to test the proposed model.
The heuristic procedure described in Section \ref{cap:cap3:sec:euristico} suggested to use the CN model with probability $0.98$, and the $VC$ and $CN$ models both with probability $0.01$. Hence, we fit a model with constant sender effects, using the MCMC algorithm described in section \ref{cap:cap3:sec:MCMC}, where the number of iterations and the hyper parameters are fixed as in the simulation setting (Section \ref{cap:cap3:sec:simulation}). We set $p=2$, the dimension of the latent space, both for plotting reasons and to compare the estimated coordinates with the geographical ones.\\
Figure \ref{cap:cap3:fig:latPosFruttaNZ}(a) reports the estimated sender parameters and the observed mean out-degrees, for the 64 countries considered. Countries corresponding to high estimated sender effects may be considered as ``top exporters''. ``Top exporting'' countries are to be interpreted as those countries that tend to export fruit to a large group of trading partners, conditionally on the the latent distances to other nodes. Three out of the first four ``top exporting'' countries are European: Italy, Netherland and Spain. Netherlands (NLD) is one of the major trade hubs for fresh fruits, importing goods from developing countries and then reselling them (mostly) to the European market \citep{cbi}\citep{olanda}. Contrary to Netherlands, Spain and Italy directly grow most of the fruits they export. Just to give an example, Canary Islands are great pineapples producers. Also, Spain in 2017 became the world's largest watermelon exporting country \citep{spagnacoco}. These three European countries show good agreement between their estimated sender effects and observed mean out-degrees, see Figure\ref{cap:cap3:fig:latPosFruttaNZ}(a). On the contrary, New Zeland, which is estimated to be the third ``top exporter'', has a moderate mean out-degree. Such mismatch between New Zeland's estimated sender effect and mean-out degree may be explained looking at its position in the latent space, in Figure \ref{cap:cap3:fig:latPosFruttaNZ}(b). Indeed, New Zeland is estimated to be quite far from the countries it trades with, therefore the ``residual'' contribution given by the sender effect to the edge probability needs to be really high. In other words, a large value of the sender parameter accounts for the ``distance'' New Zeland has to travel to export its goods.
The estimated latent coordinates for the 64 countries are also presented in Figure \ref{cap:cap3:fig:latPosFrutta}, which additionally displays the geographic coordinates of such countries, for comparison.  
The estimated latent coordinates do not resemble much the geographical ones; indeed, the Procrustes correlation value between the two sets is not high ($0.53$). The estimated latent space is characterized by a large number of Asiatic countries placed in the right side of the space and the majority of the European countries in the left part. United States are placed in between Asiatic and Oceanic countries. Indeed, most of these countries are leading suppliers of fruits for the United States, thanks to established or pending free trade agreements \citep{usa}.
The latent space is estimated to be always relevant in the determination of the edge probabilities, and to have similar effect on the different networks. Indeed, Table \ref{cap:cap3:table:alphabetaF} shows that the estimated coefficients $\bk$ range in $(0.76, 2.52)$. The same Table displays the estimates of the intercepts in the different networks, with all of the intercepts corresponding to high edge probability values (greater than $0.9$).
Figure \ref{cap:cap3:fig:probFrutta} represents the estimated probabilities in the fruit networks, given some values of the distances and of the $\theta_{i}$ parameters. In particular, the probabilities are computed for the first ($0.385$), second ($1.274$) and third ($2.281$) quartiles of the distances, and for $\theta_{i} = (-0.5, 0, 0.5, 1)$. The plot in Figure \ref{cap:cap3:fig:probFrutta} shows that, even though the latent space is constant, quite different values of the edge probability correspond to the same estimated distances, depending on the sender and the network-specific parameters, $\ak$ and $\bk$.

\begin{center}
\begin{tabular}{@{}lllllllllllll@{}}
\toprule
       & & $k=1$ & $k=2$ & $k=3$ & $k=4$ & $k=5$ & $k=6$ & $k=7$ & $k=8$ & $k=9$ & $k=10$  \\   
 \multirow{4}{*}{$\alpha$  }&        &  &  &  &  & &  &  &  &  &  & \\    
& mean &  2.00 & 1.96 & 2.51 & 2.49 & 2.63 & 1.80 & 2.26 & 2.43 & 1.78 & 2.59  \\
& sd  & - & 0.12 & 0.13 & 0.15 &  0.12 & 0.12 & 0.13 & 0.14 & 0.15 & 0.14\\
\multirow{4}{*}{$\beta$ } &     &  &  &  &  & &  &  &  &  &  \\
& mean & 1.00 & 1.37 & 0.76 & 1.04 & 1.12 & 1.70 & 1.46 & 1.13 & 2.52 & 1.42  \\
& sd   & - & 0.06 & 0.05 & 0.05 & 0.07 & 0.08 & 0.07 & 0.06 & 0.12 & 0.07 \\
\bottomrule
\end{tabular}
\captionof{table}{Averages and standard deviations of the estimated posterior distributions for the intercept and the scale coefficient parameters in the fruit networks.}
\label{cap:cap3:table:alphabetaF}
\end{center}  

\section{Discussion}
\label{cap:cap3:sec:disc}
In the present work we have introduced a novel class of Euclidean distance latent space models for multidimensional network data. The models allow to represent transitivity in a parsimonious way, via a single latent space.
Also, different levels of node-specific degree heterogeneity can be specified. In the spirit of model parsimony, we assume that the type of sender/receiver effect (``Null'', ``Constant'' and ``Variable'') is constant across the views. An interesting relaxation of such hypothesis would be to have the type of effect varying with the networks. Indeed, for example, it may be that a subset $K^*$ of the $K$ views has no sender effect, but the remaining networks have constant sender effect. Finding such sub-groups of networks would then become a clustering problem, with extra complexity brought by the allocation of each network to the specific effect-sub-group and the estimation of the number of clusters, which, however, would be bounded in $(1,3)$. \\
Also, we have proposed an heuristic procedure for model selection, that allows to choose an appropriate model for observed multiplex data without the need to estimate all the possible models. Thus, the procedure permits to bypass a classical model selection step. A preventive selection of the model may be convenient in many real data applications, as model estimation for network data can be quite (computationally) demanding. The performance of the proposed heuristic procedure and that of the latent space model have been tested in separate simulation studies and have proven to give quite good results. \\
An illustrative application to FAO trade data regarding different fruit trades has been presented, where our method was able to uncover trade patterns and shared similarities among different fruit markets. The data may be an interesting research problem per se, and an interesting extension of the proposed class of models could take into consideration weighted multiplexes, analysing import/export values or quantities. However, considering such weighted edges is non trivial, as the distributions of the exchanged quantities are both right skewed and zero-inflated.\\
The proposed models and the heuristic model selection procedure are incorporated  in the \textit{R} package \emph{spaceNet} \citep{spaceNet}, available on \emph{CRAN}.

\bibliographystyle{agsm}
\bibliography{bibliography}

\clearpage
\appendix

\section{FAO data}
\label{cap:appendice2:app1}
Below, we report the country ISO3 codes for the FAO dataset.
\begin{table}[!h]
\centering
\small
\scalebox{.95}{
\begin{tabular}{@{}lclc@{}}
\toprule
Country name        & iso3 code &Country name        & iso3 code  \\
\midrule
Australia & AUS & Malaysia& MYS \\
Austria & AUT & Maldives & MDV \\
Bahrain & BHR & Montenegro & MNE \\
Belarus & BLR & Morocco & MAR\\
Belgium & BEL & Netherlands & NLD\\                              
Bosnia and Herzegovina & BIH  & New Zealand & NZL\\
Bulgaria & BGR & Norway & NOR \\
Canada & CAN & Oman & OMN\\
Hong Kong & HKG & Pakistan & PAK \\
China & CHN & Poland & POL\\
Croatia & HRV & Portugal & PRT\\
Czech Republic & CZE & Qatar & QAT\\
Denmark & DNK & Republic of Korea & KOR \\
Egypt & EGY & Republic of Moldova & MDA\\
Estonia & EST & Romania & ROU \\
Finland & FIN & Russian Federation & RUS \\
France & FRA & Saudi Arabia & SAU \\
Germany & DEU & Serbia & SRB\\
Greece & GRC & Singapore & SGP \\
Hungary & HUN   &  Slovakia & SVK \\
India & IND & Slovenia & SVN\\
Iran & IRN & South Africa & ZAF \\
Ireland & IRL & Spain & ESP \\
Italy & ITA & Sweden & SWE \\
Jordan & JOR & Switzerland &  CHE\\
Kazakhstan & KAZ & Thailand & THA\\
Kenya & KEN &  Republic of Macedonia & MKD\\
Kuwait & KWT & Turkey & TUR \\
Latvia & LVA & Ukraine & UKR\\
Lebanon & LBN & United Arab Emirates & ARE\\
Lithuania & LTU  & United Kingdom & GBR \\
Luxembourg & LUX &  United States of America & USA \\
\bottomrule
\end{tabular}
}
\caption{Fao data: country ISO3 codes.}
\label{cap:appendice2:tab:nomiPaesiapp}
\end{table}

\begin{center}
    {{\includegraphics[width=14cm]{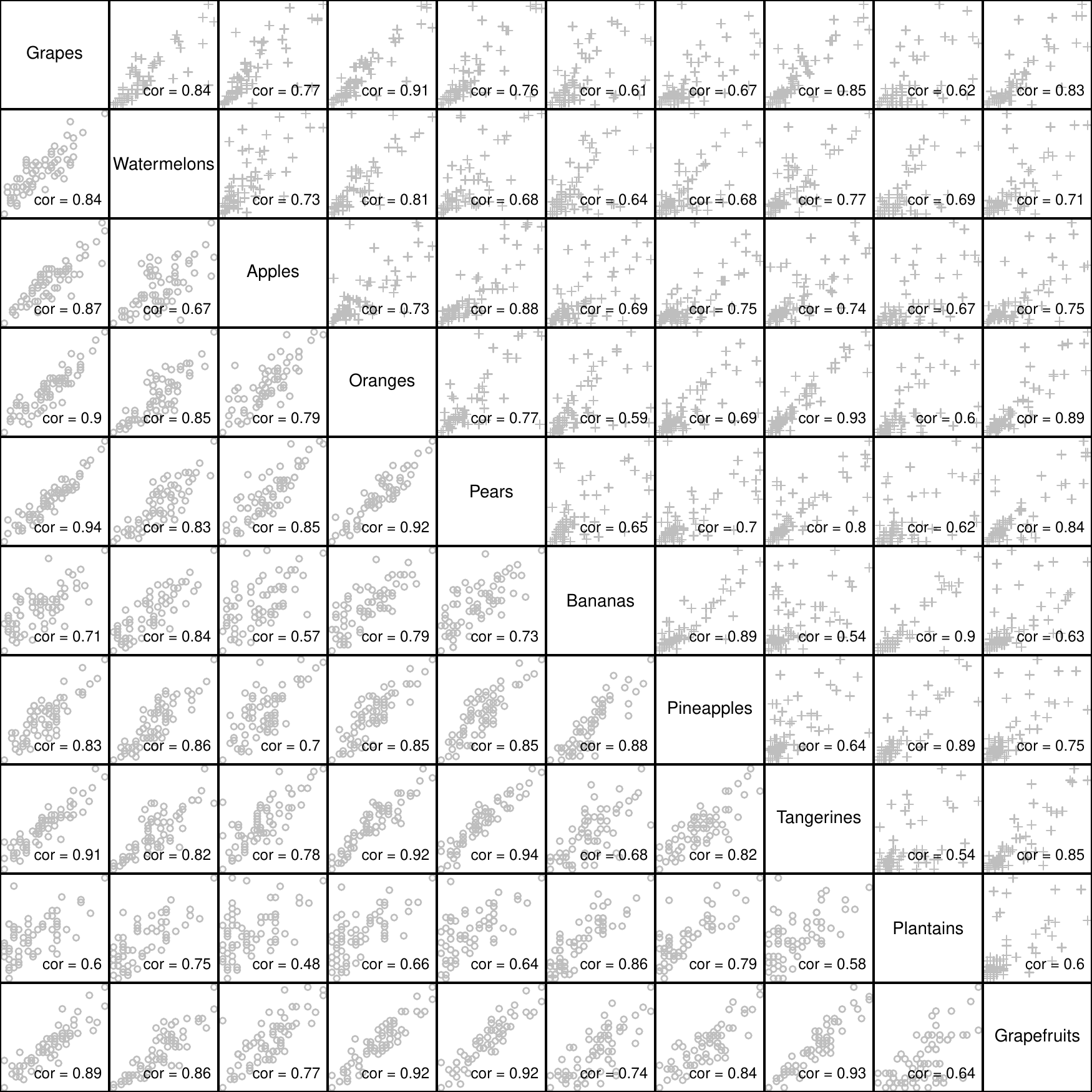} }}%
    \captionof{figure}{Fruit multiplex: Pairplots of the observed out- and in-degrees of the networks. The upper diagonal matrix represents the associations between the observed out-degrees in any couple of networks, while the lower diagonal refers to the association between the in-degrees. The values of the Spearman correlation indexes are reported.}
\end{center} 

\section{Estimation: proposal and full conditional distributions}
\label{cap:appendice2:app2}
The log-posterior distribution for the eight models with sender and/or receiver effect, presented in section \ref{cap:cap3:sec:modello}, is proportional to:
\begin{equation}
\label{cap:appendice2:eq:logposterior}
\begin{split}
 \log\Bigl( & P \bigl( \alpha, \beta,  \theta, \gamma,\mathbf{z},\mu_{\alpha}, \mu_{\beta}, \sigma_{\alpha}^2 ,\sigma_{\beta}^2 | \mathbf{Y} \bigr) \Bigr) \propto  \\
&\sum_{k = 1}^K \sum_{i = 1}^n  \sum_{j \neq i} h_{ij}^{(k)} \Bigl[ \arch \bigl(\ak \phi_{ij}^{(k)}  -\bk  d_{ij} -\sum_{f=1}^F \lambda_f x_{ijf} \bigr) - \log \Bigl( 1 + \exp\bigl\{ \ak \phi_{ij}^{(k)} 
 -\bk d_{ij} -\sum_{f=1}^F \lambda_f x_{ijf} \bigr\} \Bigr) \Bigl] \\
&-\frac{1}{2} \Biggl\{ \sum_{i = 1}^n z_i^2 + \frac{\sum_{k=1}^K (\ak -\mu_{\alpha})^2}{\sigma_{\alpha}^2} + \frac{\sum_{k=1}^K (\bk -\mu_{\beta})^2}{\sigma_{\beta}^2} 
 +K \log(\sigma_{\alpha}^2 )  +K \log(\sigma_{\beta}^2 ) + \log(\tau_{\alpha}\sigma_{\alpha}^2 )+\log(\tau_{\beta}\sigma_{\beta}^2 )\\ 
 & +\frac{\mu_{\alpha}^2}{\tau_{\alpha}\sigma_{\alpha}^2} +\frac{\mu_{\beta}^2}{\tau_{\beta}\sigma_{\beta}^2} +\frac{1}{\sigma_{\alpha}^2} + \frac{1}{\sigma_{\beta}^2} 
  + \sum_{f=1}^F \Biggl(  \frac{(\lambda_f -\mu_{\lambda_f})^2}{\sigma_{\lambda_f}^2} +\log(\sigma_{\lambda_f}^2) +\log(\tau_{\lambda}\sigma_{\lambda_f}^2) +\frac{(\mu_{\lambda_f} -\mu_{\lambda})^2}{\tau_{\lambda}\sigma_{\lambda_f}^2}  + \frac{1}{\sigma_{\lambda_f}^2}\Biggr)\Biggr\}\\
&+ \Bigl( -\frac{\nu_{\alpha} }{2} -1\Bigr)\log(\sigma_{\alpha}^2) + \Bigl( -\frac{\nu_{\beta} }{2} -1\Bigr)\log(\sigma_{\beta}^2) +\sum_{f=1}^F \Bigl(-\frac{\nu_{\lambda_f}}{2} -1\Bigr)\log(\sigma_{\lambda_f}^2),
\end{split}
\end{equation}
where, without any loss of information, the latent coordinates are assumed to be univariate.

\subsection{Nuisance parameters}
The variances of the intercept and coefficient parameters have Inverse Gamma full conditional distributions
\[
\sigma_{\alpha}^2 | \alpha, \mu_{\alpha}, \tau_{\alpha}, \nu_{\alpha}, K \thicksim \text{\small{Inv}} \Gamma\bigl( r_{\alpha}, R_{\alpha} \bigr); \qquad
\sigma_{\beta}^2 | \beta, \mu_{\beta}, \tau_{\beta}, \nu_{\beta} , K\thicksim \text{\small{Inv}} \Gamma\bigl( r_{\beta}, R_{\beta} \bigr),
\]
with parameters:
\[
r_x = \frac{\nu_x +K +1}{2}, \qquad R_x = \frac{\tau_x + \tau_x \sum_{k =1}^K (x^{(k)} -\mu_x)^2 +\mu_x^2}{2 \tau_x}.
\]
The nuisance parameters $\mu_{\alpha}, \mu_{\beta}$ are distributed as truncated normal distributions:
\[
\mu_{\alpha} | \alpha , \sigma_{\alpha}^2, \tau_{\alpha}, m_{\alpha}, K \thicksim N_{\bigl[0, \infty \bigl]} \Biggl( \frac{\tau_{\alpha} \sum_{k =1}^K \ak + m_{\alpha}}{1 + K\tau_{\alpha} }, \frac{\tau_{\alpha}\sigma_{\alpha}^2}{1 + K\tau_{\alpha}}\Biggr)\]
\[
\mu_{\beta} | \beta , \sigma_{\beta}^2, \tau_{\beta},m_{\beta},K \thicksim N_{\bigr[0, \infty \bigl]} \Biggl( \frac{\tau_{\beta} \sum_{k =1}^K \bk +m_{\beta}}{1 + K\tau_{\beta} }, \frac{\tau_{\beta}\sigma_{\beta}^2}{1 + K\tau_{\beta}}\Biggr) .
\]
  
\subsection{Latent positions}
\label{cap:appendice2:lat_pos_prop}
The proposal distribution for the $i^{\text{th}}$ latent coordinate is derived from the log-posterior distribution of the model in equation \ref{cap:appendice2:eq:logposterior}. The logarithmic term in the log-likelihood, which is indeed a LSE function, is approximated with its lower bound.
\[
\tilde{z}_i \mid \mathbf{Y},\alpha, \ti, \gamma, \beta, \lambda, \mathbf{x}_i,\mathbf{D}, K \thicksim N \Bigl(\mu_{\tilde{z}_i}, \sigma_{\tilde{z}_i}^2 \Bigr),
\]
where 
\[
\mu_{\tilde{z}_i} = \sigma_{\tilde{z}_i}^2 \Bigl(2 \sum_{k = 1}^K \bk \sum_{j \neq i} h_{ij}^{(k)} \bigl(\arch -w_{ij}^{(k)} \bigr) z_j \Bigr), \quad 
 \sigma_{\tilde{z}_i}^2  = \Biggl( 1 + 2 \sum_{k = 1}^K  \bk \sum_{j \neq i} h_{ij}^{(k)} |\arch -w_{ij}^{(k)}| \Biggr)^{-1}.
\]
where $w_{ij}^{(k)}$ is a binary indicator variable, defined as:
\[
w_{ij}^{(k)}= 
\begin{cases}
1 \qquad \text{if} \quad \ak 	\phi_{ij}^{(k)}  -\bk d_{ij} -\sum_{f=1}^F \lambda_f x_{ijf}  >0  \\
0 \qquad \text{if} \quad \ak \phi_{ij}^{(k)}   -\bk d_{ij} -\sum_{f=1}^F \lambda_f x_{ijf}  \leq 0
   \end{cases}
\]

\subsection{Intercept parameters}
The proposal distribution for the $k^{\text{th}}$ intercept parameter $\alpha^{(k)}$ is derived from the log-posterior distribution, where the logarithmic term is approximated via its second order Taylor expansion in $\alpha^{(k)} = \mu_{\alpha}$.
Defining $E^{(k)} = \sum_{i = 1}^n  \sum_{j \neq i} h_{ij}^{(k)} \arch \phi_{ij}^{(k)} $, the proposal distribution for intercept $\alpha^{(k)}$ is taken to be:
\[
\tilde{\alpha}^{(k)} \mid \mathbf{Y}, \mathbf{D}, \mathbf{H}, \bk, \theta^{(k)},\gamma^{(k)},\lambda, \mathbf{X} ,\mu_{\alpha}, \sigma_{\alpha}^2 \thicksim N \Bigl(\mu_{\tilde{\alpha}^{(k)}}, \sigma_{\tilde{\alpha}^{(k)}}^2 \Bigr),
\]
with
\[
\mu_{\tilde{\alpha}^{(k)}} = \sigma_{\tilde{\alpha}^{(k)}}^2 \Biggl\{ E^{(k)} -  \sum_{i = 1}^n  \sum_{j \neq i}  \frac{h_{ij}^{(k)} \phi_{ij}^{(k)}  \exp\bigl\{ \mu_{\alpha} \phi_{ij}^{(k)}  -\bk d_{ij} -\sum_{f=1}^F \lambda_f x_{ijf} \bigr\} }{1 + \exp\bigl\{ \mu_{\alpha}\phi_{ij}^{(k)}  -\bk d_{ij} -\sum_{f=1}^F \lambda_f x_{ijf}  \bigr\}} \Biggr\} +\mu_{\alpha},
\]
\[
\sigma_{\tilde{\alpha}^{(k)}}^2 = \Biggl\{ \sum_{i = 1}^n  \sum_{j \neq i} \frac{ h_{ij}^{(k)}( \phi_{ij}^{(k)} )^2\exp\bigl\{ \mu_{\alpha} \phi_{ij}^{(k)} -\bk d_{ij} -\sum_{f=1}^F \lambda_f x_{ijf} \bigr\} }{\bigl( 1 + \exp\bigl\{ \mu_{\alpha}\phi_{ij}^{(k)}  -\bk d_{ij}-\sum_{f=1}^F \lambda_f x_{ijf}  \bigr\}\bigr)^2}  +\frac{1}{\sigma_{\alpha}^2}\Biggr\}^{-1}.
\]

\subsection{Coefficient parameters (distances)}
The proposal distribution for the $k^{\text{th}}$ coefficient parameter $\beta^{(k)}$ is derived from the log-posterior distribution, where the logarithmic term is approximated via its second order Taylor expansion in $\beta^{(k)} = \mu_{\beta}$.
Then, the proposal distribution specified for intercept coefficient $k$ is:
\[
\tilde{\beta}^{(k)} \mid \mathbf{Y}, \mathbf{D}, \mathbf{H} ,\ak, \theta^{(k)},\gamma^{(k)},\lambda, \mathbf{X} ,\mu_{\beta}, \sigma_{\beta}^2 \thicksim N \Bigl(\mu_{\tilde{\beta}^{(k)}}, \sigma_{\tilde{\beta}^{(k)}}^2,n \Bigr) ,
\]
where
\[
 \mu_{\tilde{\beta}^{(k)}} = \sigma_{\tilde{\beta}^{(k)}}^2 \Biggl\{  \sum_{i = 1}^n  \sum_{j \neq i} h_{ij}^{(k)} d_{ij} \Biggl( \frac{\exp\bigl\{ \ak \phi_{ij}^{(k)} -\mu_{\beta} d_{ij} -\sum_{f=1}^F \lambda_f x_{ijf} \bigr\} }{1 + \exp\bigl\{ \ak \phi_{ij}^{(k)} -\mu_{\beta} d_{ij} -\sum_{f=1}^F \lambda_f x_{ijf} \bigr\}}  -\arch  \Biggr) \Biggr\} +\mu_{\beta},
\]
\[
 \sigma_{\tilde{\beta}^{(k)}}^2 = \Biggl\{ \sum_{i = 1}^n  \sum_{j \neq i} \frac{ h_{ij}^{(k)}  d_{ij}^2 \exp\bigl\{ \ak\phi_{ij}^{(k)}  -\mu_{\beta} d_{ij} -\sum_{f=1}^F \lambda_f x_{ijf}  \bigr\} }{\bigl( 1 + \exp\bigl\{ \ak \phi_{ij}^{(k)}-\mu_{\beta} d_{ij} -\sum_{f=1}^F \lambda_f x_{ijf}\bigr\}\bigr)^2}  +\frac{1}{\sigma_{\beta}^2}\Biggr\}^{-1}.
\]
\subsection{Coefficient parameters (covariates)} 
The proposal distribution for $\lambda_f$ is:
\[
\tilde{\lambda}_f \mid \alpha, \beta, \lambda, \gamma, \theta, \mu_{\lambda_{f}},\sigma_{\lambda_f}^2 ,\mathbf{D},\mathbf{H}, \mathbf{X}, \mathbf{Y} \thicksim N_{(0, \infty)}\bigl( \mu_{\tilde{\lambda}_f}, \sigma_{\tilde{\lambda}_f}^2 \bigr),
\]
where 
\[
\mu_{\tilde{\lambda}_f} = \sigma_{\tilde{\lambda}_f}^2\Biggl[ \sum_{k = 1}^K \Biggl( \sum_{i = 1}^n \sum_{j \neq i}  \frac{ x_{ijf} h_{ij}^{(k)} \exp \bigl( \ak \phi_{ij}^{(k)}-\bk d_{ij} -\sum_{l \neq f} \lambda_l x_{ijl} \bigr)}{1 +\exp \bigl( \ak \phi_{ij}^{(k)}-\bk d_{ij} -\sum_{l \neq f} \lambda_l x_{ijl} \bigr)} -y_{ij}^{(k)} x_{ijf} \Biggr)
\Biggr] +\mu_{\lambda_f},
\]
\[
\sigma_{\tilde{\lambda}_f}^2 =  \Biggl\{ \sum_{k = 1}^K \sum_{i = 1}^n \sum_{j \neq i}  \frac{h_{ij}^{(k)} x_{ijf}^2 \exp \bigl( \ak\phi_{ij}^{(k)}-\bk d_{ij} -\sum_{l \neq f} \lambda_l x_{ijl} \bigr)}{\Bigl(1 +\exp \bigl( \ak \phi_{ij}^{(k)} -\bk d_{ij} -\sum_{l \neq f} \lambda_l x_{ijl} \bigr) \Bigr)^2} +\frac{1}{\sigma_{\lambda_f}^2} \Biggr\}^{-1}.
\]
\subsection{Sender and receiver parameters}
To define proposal distributions for the sender and receiver parameter, we consider only the relevant parts of the posterior distribution. In particular, we derive them from the log-likelihood of the model, that is, the first line of the log-posterior distribution presented in equation \ref{cap:appendice2:eq:logposterior},
\[
\sum_{k = 1}^K \sum_{i = 1}^n  \sum_{j \neq i} h_{ij}^{(k)} \Bigl[ \arch \bigl(\ak \phi_{ij}^{(k)}  -\bk  d_{ij} -\sum_{f=1}^F \lambda_f x_{ijf} \bigr) - \log \Bigl( 1 + \exp\bigl\{ \ak \phi_{ij}^{(k)}  
-\bk d_{ij} -\sum_{f=1}^F \lambda_f x_{ijf} \bigr\} \Bigr) \Bigl].
\]
\subsubsection{Sender parameters}
When the sender effect is variable across the networks, we need to propose a new value for the $i^{\text(th)}$ sender parameter in the $k^{\text{th}}$ network at the $t^{\text{th}}$ iteration of the MCMC algorithm. We start by  approximating the logarithmic term in the log-likelihood via its second order Taylor's expansion in  $\theta_{i}^{(k)t}\approx \theta_{i}^{(k)t-1}$:
\[
\begin{split}
\log \Bigl( 1 + \exp\bigl\{ \ak &\phi_{ij}^{(k)t}  -\bk d_{ij} \bigr\} \Bigr) = \log \Bigl( 1 + \exp\bigl\{  \ak \theta_{i}^{(k)t}x + \ak \gamma_{j}^{(k)}x -\bk d_{ij} \bigr\} \Bigr)\\
& \approx \Bigl( \theta_{i}^{(k)t} -\theta_{i}^{(k)t-1} \Bigr) \Biggl\{ \frac{\ak x \exp\bigl\{  \ak \theta_{i}^{(k)t-1}x + \ak \gamma_{j}^{(k)}x  -\bk d_{ij} \bigr\} }{1 + \exp\bigl\{ \ak \theta_{i}^{(k)t-1}x + \ak \gamma_{j}^{(k)}x  -\bk d_{ij} \bigr\} } \Biggr\}
+\frac{1}{2}\Bigl( \theta_{i}^{(k)t} -\theta_{i}^{(k)t-1} \Bigr)^2 \\
&\Biggl\{ \frac{ \bigl(\ak x\bigr)^2 \exp\bigl\{ \ak \theta_{i}^{(k)t-1}x + \ak \gamma_{j}^{(k)}x  -\bk d_{ij} \bigr\} }{\bigl( 1 + \exp\bigl\{\ak \theta_{i}^{(k)t-1}x+ \ak \gamma_{j}^{(k)}x  -\bk d_{ij} \bigr\} \bigr)^2} \Biggr\} +const\\
\end{split},
\]
with $x=1$ when only one type of effect is present and $x=0.5$ when both types are.\\
Then, the logarithmic term is substituted with its approximation. The approximated log-likelihood is now a quadratic function in $\theta_i^{(k)t}$, which permits to define the following proposal distribution
\[
\tilde{\theta}_i^{(k)t} \mid \ak, \bk, \lambda, \gamma^{(k)}, \theta_i^{(k)t-1}, \mathbf{D},\mathbf{H}, \mathbf{X}, \mathbf{Y} \thicksim N_{(-1,1)}\Bigl( \mu_{\tilde{\theta}_i^{(k)t}} ,  \sigma_{\tilde{\theta}_i^{(k)t}}^2 \Bigr)
\]
where
\[
 \mu_{\tilde{\theta}_i^{(k)t}}  = \Biggl[ \alpha^{(k)}x \sum_{j \neq i} h_{ij}^{(k)}  \Biggl(      y_{ij}^{(k)} - \frac{\exp \bigl(\alpha^{(k)}\phi_{ij}^{(k)t-1}  -\beta^{(k)} d_{ij} \bigr) }{1 + \exp \bigl(\alpha^{(k)}\phi_{ij}^{(k)t-1}  -\beta^{(k)} d_{ij} \bigr)}\Biggr)\Biggr]\sigma_{\tilde{\theta}_i^{(k)t}}^2 + \theta_{i}^{(k)t-1}
\]

\[
\sigma_{\tilde{\theta}_i^{(k)t}}^2  = \Biggl[ \Biggl(\alpha^{(k)}x \Biggr)^2 \sum_{j \neq i} h_{ij}^{(k)} \Biggl(\frac{\exp \bigl(\alpha^{(k)}\phi_{ij}^{(k)t-1}  -\beta^{(k)} d_{ij} \bigr) }{\Bigl( 1 + \exp \bigl(\alpha^{(k)}\phi_{ij}^{(k)t-1}  -\beta^{(k)} d_{ij} \bigr)\Bigr)^2}\Biggr)\Biggr]^{-1}
\]
Instead, when the sender effect is constant, the logarithmic term in the log-likelihood is approximated with its second order Taylor's expansion in  $\theta_{i}^{t}\approx \theta_{i}^{t-1}$a and the resulting proposal distribution is:               
   \[
\tilde{\theta}_i^{t} \mid \alpha, \beta, \lambda, \gamma, \theta_i^{t-1}, \mathbf{D},\mathbf{H}, \mathbf{X}, \mathbf{Y}\thicksim N_{(-1,1)}\Bigl( \mu_{\tilde{\theta}_i^{t}} ,  \sigma_{\tilde{\theta}_i^{t}}^2 \Bigr)
\]
where
\[
 \mu_{\tilde{\theta}_i^{t}}  = \Biggl[  \sum_{k=1}^K \alpha^{(k)}x \sum_{j \neq i}h_{ij}^{(k)}  \Biggl(      y_{ij}^{(k)} - \frac{\exp \bigl(\alpha^{(k)}\phi_{ij}^{t-1}  -\beta^{(k)} d_{ij} \bigr) }{1 + \exp \bigl(\alpha^{(k)}\phi_{ij}^{(t-1}  -\beta^{(k)} d_{ij} \bigr)}\Biggr)\Biggr]\sigma_{\tilde{\theta}_i^{t}}^2 + \theta_{i}^{t-1}
\]

\[
\sigma_{\tilde{\theta}_i^{t}}^2  = \Biggl[ \sum_{k=1}^K \Biggl(\alpha^{(k)}x \Biggr)^2 \sum_{j \neq i}h_{ij}^{(k)}  \Biggl(\frac{\exp \bigl(\alpha^{(k)}\phi_{ij}^{t-1}  -\beta^{(k)} d_{ij} \bigr) }{\Bigl( 1 + \exp \bigl(\alpha^{(k)}\phi_{ij}^{t-1}  -\beta^{(k)} d_{ij} \bigr)\Bigr)^2}\Biggr)\Biggr]^{-1}
\]                               
The proposal distribution defined depend on the previous value of the parameter of interest, as in random walk proposals, but also on the current configuration of all the other parameters. This allows to explore efficiently the parameter space for the $\theta_i^{(k)}$s. Indeed, classical random walks are usually slow in the exploration of the parameter space and, when employed in the estimation procedure of this class of models, they returned unstable chains. ``Correcting'' the random walk by incorporating the past value of the sender parameter of interest in a more complex parametrization of the proposal distribution has proven to be a valid approach to estimate these parameters. 
\subsubsection{Receiver parameters}
The proposal distributions for the receiver parameters, both in the variable and in the constant case, are recovered with the same procedure used to define those of the sender parameters. The only, trivial, difference is that the logarithmic term of the log-likelihood is approximated in the receiver parameter.
Then, the proposal distributions for the $j^{\text{th}}$ receiver parameter at the $t^{\text{th}}$ iteration are, if the effect is variable:
\[
\tilde{\gamma}_j^{(k)t} \mid \ak, \bk, \lambda, \theta^{(k)}, \gamma_j^{(k)t-1}, \mathbf{D},\mathbf{H}, \mathbf{X}, \mathbf{Y}\thicksim N_{(-1,1)}\Bigl( \mu_{\tilde{\gamma}_j^{(k)t}} ,  \sigma_{\tilde{\gamma}_j^{(k)t}}^2 \Bigr)
\]
where
\[
 \mu_{\tilde{\gamma}_j^{(k)t}}  = \Biggl[ \alpha^{(k)}x \sum_{i \neq j}h_{ij}^{(k)} \Biggl(y_{ij}^{(k)} - \frac{\exp \bigl(\alpha^{(k)}\phi_{ij}^{(k)t-1}  -\beta^{(k)} d_{ij} \bigr) }{1 + \exp \bigl(\alpha^{(k)}\phi_{ij}^{(k)t-1}  -\beta^{(k)} d_{ij} \bigr)}\Biggr)\Biggr]\sigma_{\tilde{\gamma}_j^{(k)t}}^2 + \gamma_{j}^{(k)t-1}
\]
\[
\sigma_{\tilde{\gamma}_j^{(k)t}}^2  = \Biggl[ \Biggl(\alpha^{(k)}x \Biggr)^2 \sum_{i \neq j} h_{ij}^{(k)} \Biggl(\frac{\exp \bigl(\alpha^{(k)}\phi_{ij}^{(k)t-1}  -\beta^{(k)} d_{ij} \bigr) }{\Bigl( 1 + \exp \bigl(\alpha^{(k)}\phi_{ij}^{(k)t-1}  -\beta^{(k)} d_{ij} \bigr)\Bigr)^2}\Biggr)\Biggr]^{-1},
\]
and if the effect is constant:
 \[
\tilde{\gamma}_j^{t} \mid \alpha, \beta, \lambda, \theta, \gamma_j^{t-1}, \mathbf{D},\mathbf{H}, \mathbf{X}, \mathbf{Y}\thicksim N_{(-1,1)}\Bigl( \mu_{\tilde{\gamma}_j^{t}} ,  \sigma_{\tilde{\gamma}_j^{t}}^2 \Bigr)
\]
where
\[
 \mu_{\tilde{\gamma}_j^{t}}  = \Biggl[  \sum_{k=1}^K \alpha^{(k)}x \sum_{i \neq j} h_{ij}^{(k)} \Biggl(      y_{ij}^{(k)} - \frac{\exp \bigl(\alpha^{(k)}\phi_{ij}^{t-1}  -\beta^{(k)} d_{ij} \bigr) }{1 + \exp \bigl(\alpha^{(k)}\phi_{ij}^{(t-1}  -\beta^{(k)} d_{ij} \bigr)}\Biggr)\Biggr]\sigma_{\tilde{\gamma}_j^{t}}^2 + \gamma_{j}^{t-1}
\]
\[
\sigma_{\tilde{\gamma}_j^{t}}^2  = \Biggl[ \sum_{k=1}^K \Biggl(\alpha^{(k)}x\Biggr)^2 \sum_{i \neq j}h_{ij}^{(k)}  \Biggl(\frac{\exp \bigl(\alpha^{(k)}\phi_{ij}^{t-1}  -\beta^{(k)} d_{ij} \bigr) }{\Bigl( 1 + \exp \bigl(\alpha^{(k)}\phi_{ij}^{t-1}  -\beta^{(k)} d_{ij} \bigr)\Bigr)^2}\Biggr)\Biggr]^{-1}.
\]                               
\subsubsection{Undirected networks}
The proposal distributions for the $\delta_{i}^{(k)}$ parameters, both in the variable and in the constant case, are recovered with the same procedure used to define those of the sender and the receiver parameters. This time, the logarithmic term in the log-likelihood, which this time is defined via the edge probabilities in equation \ref{cap:cap3:eq:socprob}, is approximated in the $i^{\text{th}}$ parameter.
The proposal distributions for the $i^{\text{th}}$ parameter at the $t^{\text{th}}$ iteration are, if the effect is variable:
\[
\tilde{\delta}_i^{(k)t} \mid \ak, \bk, \lambda, \delta^{(k)t-1}, \mathbf{D},\mathbf{H}, \mathbf{X}, \mathbf{Y} \thicksim N_{(-1,1)}\Bigl( \mu_{\tilde{\delta}_i^{(k)t}} ,  \sigma_{\tilde{\delta}_i^{(k)t}}^2 \Bigr)
\]
where
\[
 \mu_{\tilde{\delta}_i^{(k)t}}  = \Biggl[ \frac{\alpha^{(k)}}{2} \sum_{j \neq i} h_{ij}^{(k)}  \Biggl(      y_{ij}^{(k)} - \frac{\exp \bigl(\alpha^{(k)}\phi_{ij}^{(k)t-1}  -\beta^{(k)} d_{ij} \bigr) }{1 + \exp \bigl(\alpha^{(k)}\phi_{ij}^{(k)t-1}  -\beta^{(k)} d_{ij} \bigr)}\Biggr)\Biggr]\sigma_{\tilde{\delta}_i^{(k)t}}^2 + \delta_{i}^{(k)t-1}
\]
\[
\sigma_{\tilde{\delta}_i^{(k)t}}^2  = \Biggl[ \Biggl(\frac{\alpha^{(k)}}{2} \Biggr)^2 \sum_{j \neq i}h_{ij}^{(k)}  \Biggl(\frac{\exp \bigl(\alpha^{(k)}\phi_{ij}^{(k)t-1}  -\beta^{(k)} d_{ij} \bigr) }{\Bigl( 1 + \exp \bigl(\alpha^{(k)}\phi_{ij}^{(k)t-1}  -\beta^{(k)} d_{ij} \bigr)\Bigr)^2}\Biggr)\Biggr]^{-1},
\]
and if the effect is constant:
 \[
\tilde{\delta}_i^{t} \mid \alpha, \beta, \lambda, \delta^{t-1}, \mathbf{D},\mathbf{H}, \mathbf{X}, \mathbf{Y}\thicksim N_{(-1,1)}\Bigl( \mu_{\tilde{\delta}_i^{t}} ,  \sigma_{\tilde{\delta}_i^{t}}^2 \Bigr)
\]
where
\[
 \mu_{\tilde{\delta}_i^{t}}  = \Biggl[  \sum_{k=1}^K \frac{\alpha^{(k)}}{2} \sum_{j \neq i}h_{ij}^{(k)}  \Biggl(      y_{ij}^{(k)} - \frac{\exp \bigl(\alpha^{(k)}\phi_{ij}^{t-1}  -\beta^{(k)} d_{ij} \bigr) }{1 + \exp \bigl(\alpha^{(k)}\phi_{ij}^{(t-1}  -\beta^{(k)} d_{ij} \bigr)}\Biggr)\Biggr]\sigma_{\tilde{\delta}_i^{t}}^2 + \delta_{i}^{t-1}
\]
\[
\sigma_{\tilde{\delta}_i^{t}}^2  = \Biggl[ \sum_{k=1}^K \Biggl(\frac{\alpha^{(k)}}{2} \Biggr)^2 \sum_{j \neq i}h_{ij}^{(k)}  \Biggl(\frac{\exp \bigl(\alpha^{(k)}\phi_{ij}^{t-1}  -\beta^{(k)} d_{ij} \bigr) }{\Bigl( 1 + \exp \bigl(\alpha^{(k)}\phi_{ij}^{t-1}  -\beta^{(k)} d_{ij} \bigr)\Bigr)^2}\Biggr)\Biggr]^{-1}.
\]                               
    
\clearpage

\section{Scenario I: simulation results}
\label{cap:cap3:app:simRes}

\begin{figure}[!h]
    \centering
    {{\includegraphics[width=14cm]{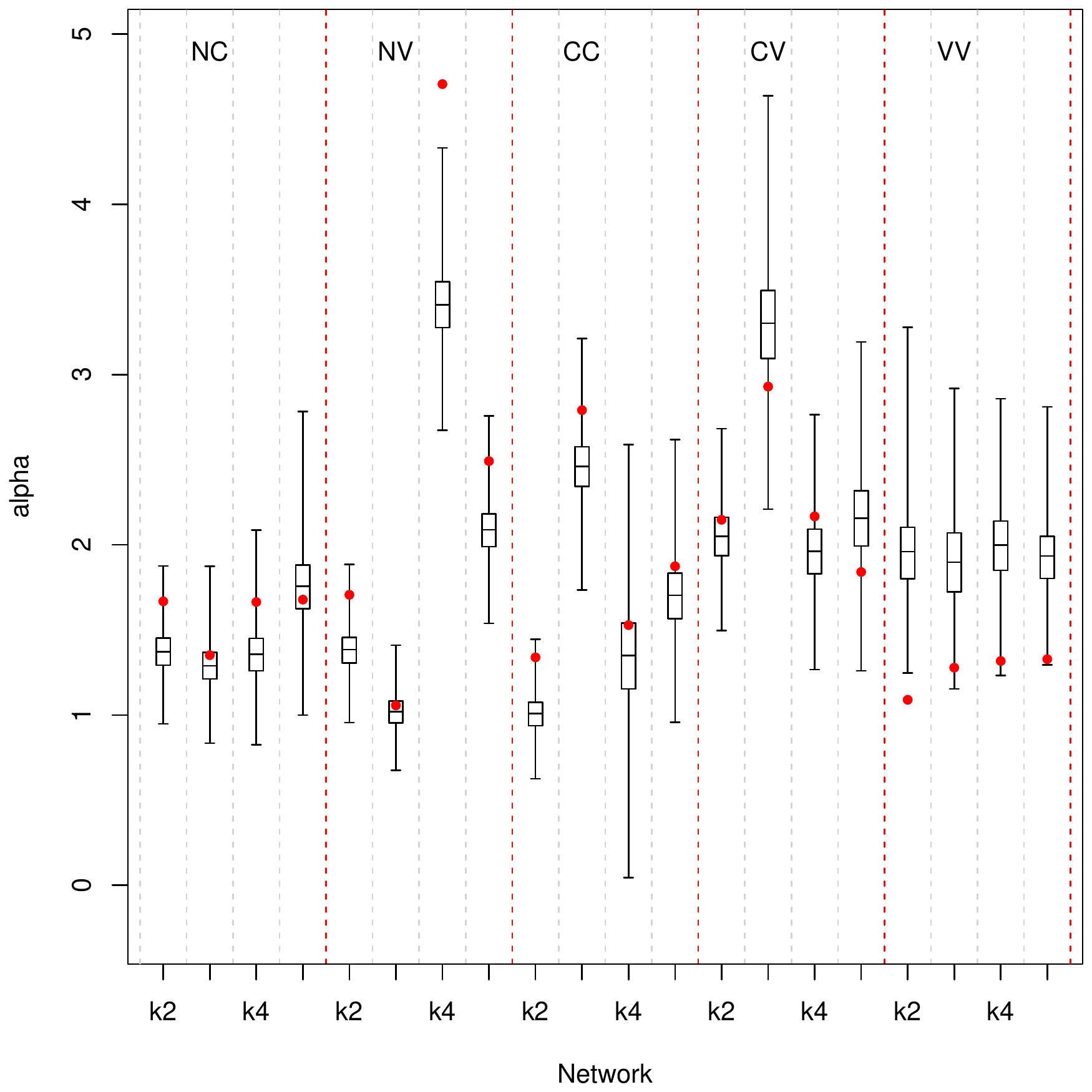} }}%
    \caption{Boxplots of the estimated posterior distributions for the intercept parameters $\ak$. Red dots indicate the true, simulated, values of the intercepts.}
    \label{cap:appendice2:fig:alphaB1sim}%
\end{figure}

\begin{figure}[!h]
    \centering
    {{\includegraphics[width=14cm]{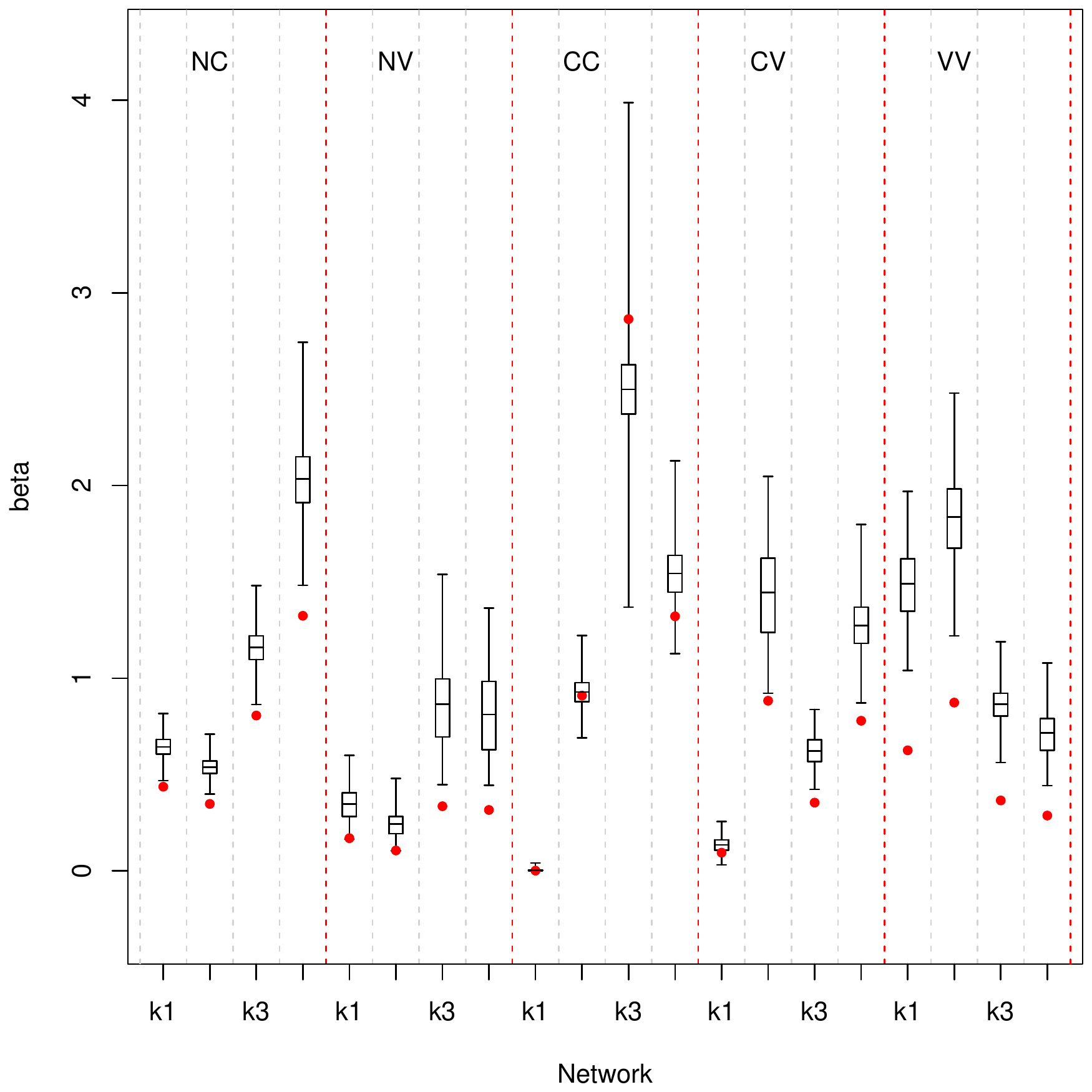} }}%
    \caption{Boxplots of the estimated posterior distributions for the coefficient parameters $\bk$. Red dots indicate the true, simulated, values of the coefficients.}
    \label{cap:appendice2:fig:betaB1sim}%
\end{figure}

\end{document}